%% file: TSE-2025-02-0177.R2_Chen.tex
\documentclass[lettersize,journal]{IEEEtran}
\usepackage{amsmath,amsfonts}
\usepackage{algorithmic}
\usepackage{algorithm}
\usepackage{array}
\usepackage[caption=false,font=normalsize,labelfont=sf,textfont=sf]{subfig}
\usepackage{textcomp}
\usepackage{stfloats}
\usepackage{url}
\usepackage{verbatim}
\usepackage{graphicx}
\usepackage{cite}
\usepackage{mycommands}
\usepackage{hyperref}
\usepackage{xcolor}
\newcommand{\revised}[1]{#1} %main manuscript
\newcommand{\revRtwo}[1]{#1}
\usepackage{subcaption}
\usepackage{graphicx}

\begin{document}

\title{Are Decoder-Only Large Language Models the Silver Bullet for Code Search?}

\author{Yuxuan~Chen,
        Mingwei~Liu,
        Guangsheng~Ou,
        Anji~Li,
        Dekun~Dai,
        Yanlin~Wang,
        and~Zibin~Zheng$^{*}$,~\IEEEmembership{Fellow,~IEEE}%
\IEEEcompsocitemizethanks{
    \IEEEcompsocthanksitem Y. Chen, M. Liu, G. Ou, A. Li, D. Dai, and Y. Wang are with the School of Software Engineering, Sun Yat-sen University, Zhuhai 519082, China, and also with the Zhuhai Key Laboratory of Trusted Large Language Models, Zhuhai, China.
    \protect\\
    E-mail: \{chenyx677, ougsh3, lianj8, daidk\}@mail2.sysu.edu.cn, \{liumw26, wangylin36\}@mail.sysu.edu.cn
    \IEEEcompsocthanksitem Z. Zheng is with the School of Software Engineering, Sun Yat-sen University, Zhuhai 519082, China, and also with the Shenzhen Research Institute of Sun Yat-sen University, Shenzhen, China.
    \protect\\
    E-mail: zhzibin@mail.sysu.edu.cn
    \IEEEcompsocthanksitem *Z. Zheng is the corresponding author.}
\thanks{This work was supported by National Natural Science Foundation of China (Grant No. 62402113), the General Program of the Natural Science Foundation of Guangdong Province, China (Grant No. 2025A1515011631), and Shenzhen Science and Technology Program (JCYJ20240813150107010).}}

\markboth{Journal of \LaTeX\ Class Files,~Vol.~14, No.~8, August~2021}%
{Chen \MakeLowercase{\textit{et al.}}: Are Decoder-Only Large Language Models the Silver Bullet for Code Search?}

\maketitle

\begin{abstract}
Code search is essential for code reuse, allowing developers to efficiently locate relevant code snippets.
\revised{The advent of powerful decoder-only Large Language Models (LLMs) has revolutionized many code intelligence tasks. However, their effectiveness for the retrieval-based task of code search, particularly compared to established encoder-based models, remains underexplored. This paper addresses this gap by presenting a large-scale systematic evaluation of eleven decoder-only LLMs, analyzing their performance across zero-shot and fine-tuned settings.}

\revised{Our results show that fine-tuned decoder-only models, particularly CodeGemma, significantly outperform encoder-only models like UniXcoder, achieving a 40.4\% higher Mean Average Precision (MAP) on the CoSQA$^+$ benchmark. Our analysis further reveals two crucial nuances for practitioners: first, the relationship between model size and performance is non-monotonic, with mid-sized models often outperforming larger variants; second, the composition of the training data is critical, as a multilingual dataset enhances generalization while a small amount of data from a specific language can act as noise and interfere with model effectiveness. These findings offer a comprehensive guide to selecting and optimizing modern LLMs for code search.}

\end{abstract}


\input{sections/introduction}

\input{sections/background}

\input{sections/study_design}

\input{sections/rq1}

\input{sections/rq2/rq2}

\input{sections/rq2/rq2-design}

\input{sections/rq2/rq2-result}

\input{sections/RQ3/RQ3}

\input{sections/RQ4/RQ4}
\input{sections/rq6}

\input{sections/RelatedWork}
\input{sections/threat}

\input{sections/Practical_Implications}

\input{sections/conclusion}

\bibliographystyle{IEEEtran}

\bibliography{reference}

\vfill

\end{document}

%% file: sections/introduction.tex
\section{Introduction}

Code search is a fundamental process in software engineering, allowing developers to retrieve and query semantically relevant code snippets from large-scale codebases using natural language queries (NL-to-Code) \cite{codesurvey}. This capability is crucial for code reuse, discovering relevant examples, and accelerating the learning and onboarding process for developers \cite{zheng2023towards}. With the advancement of large language models (LLMs) such as ChatGPT \cite{chatgpt2023} and DeepSeekCoder \cite{guo2024deepseek}, Retrieval-Augmented Generation (RAG) has become a prominent approach to enhancing their capabilities in code-related tasks \cite{gao2023retrieval}. By retrieving and incorporating relevant code snippets, LLMs can significantly improve performance in software engineering tasks like code generation, among others \cite{austin2021program}. However, the accuracy of code search remains a critical bottleneck \cite{codesurvey,shi2023cocosoda}.

The prevailing approach to code search employs a pre-training and fine-tuning paradigm \cite{guo2020graphcodebert}, typically using encoder-based models like CodeBERT \cite{feng2020codebert} or \unixcoder \cite{guo2022unixcoder}. In this method, natural language queries and code are converted into vector embeddings and fine-tuned with specific NL-to-Code datasets. This contrastive learning process helps capture semantic similarities, aligning related queries and code snippets in the vector space.

\revised{ However, these pioneering encoder-based models face a fundamental bottleneck when compared to the current generation of LLMs: a significant gap in scale and semantic capability. Models like CodeBERT, typically with hundreds of millions of parameters, are orders of magnitude smaller than modern decoder-only LLMs \cite{feng2020codebert}. This vast difference in scale, originating from billions of parameters and pre-training on trillions of tokens, endows LLMs with superior semantic understanding and reasoning abilities \cite{ahmad2021unified}. This raises a critical question: can these advanced capabilities address the inherent generalization limitations of smaller models and redefine the performance ceiling for code search?}

Decoder-only LLMs excel in natural language processing and software engineering tasks, outperforming fine-tuned small-scale models in areas like code generation and program repair \cite{chen2021evaluating}. Their ability to process \revised{long and complex contexts} \cite{rae2021scaling}, combined with rich pre-training data and large-scale parameters, enhances code and natural language understanding \cite{chen2021evaluating, zheng2023towards, zheng2023survey}, potentially addressing the bottlenecks in code search tasks. 

Despite their potential, the application of decoder-only LLMs, such as \deepseekllm \cite{bi2024deepseek}, CodeLlama \cite{roziere2023code}, and DeepSeekCoder \cite{guo2024deepseek}, to code search tasks remains underexplored. The suitability of these models for code search is still unclear. One key challenge is that code search is not a generative task like code generation or program repair, which better align with the pretraining tasks typically used for decoder-only LLMs. As a result, directly applying decoder-only LLMs to code search may not be the most effective approach. A deeper investigation is needed to determine whether these models can be effectively used for code search, how best to leverage them for this task, and whether they can outperform existing approaches based on smaller encoder-only pre-trained models. This paper aims to explore whether decoder-only LLMs could be the ``silver bullet'' for code search, offering new possibilities for improving efficiency and accuracy.

\textbf{Study Design.}
To address this gap, we systematically evaluate the performance of state-of-the-art (SOTA) decoder-only LLMs in code search.
Specifically, we investigate the following research questions (RQs):

\begin{itemize}[leftmargin=15pt]
     \item \textbf{RQ1 (Zero-shot Performance)}: How well do decoder-only LLMs perform on code search without fine-tuning?
    
     \item \textbf{RQ2 (Fine-tuning Improvement)}: 
     \revised{To what extent does fine-tuning improve the performance of decoder-only LLMs over the zero-shot setting?}

     \item \textbf{RQ3 (Improvement Analysis)}: What factors contribute to performance gains from fine-tuning?
     \begin{itemize}[leftmargin=5pt]
     
     \item \textbf{RQ3.a (Training Method)}: How do different fine-tuning strategies impact performance?
     
     \item \textbf{RQ3.b (Training Data)}: How does the quality and type of training data influence fine-tuning effectiveness?

     \item \textbf{\revised{RQ3.c (Single-Language Fine-Tuning)}}: \revised{How the specific programming language used for fine-tuning influences a model's ability to learn generalizable code representations?}
     
     \item \textbf{RQ3.d (Model Size)}: How does the model size influence the effectiveness of fine-tuning in code search tasks?  
    
    \item \textbf{RQ3.e (Query and Code Length)}: How do query and code lengths affect model performance?    \end{itemize}
    
     \item \textbf{RQ4 (Computational Time)}: How does the computational time of decoder-only LLMs compare to that of smaller encoder-only models in code search tasks?
    
    \item \textbf{RQ5 (Training Efficiency)}: How efficiently do decoder-only LLMs learn and generalize during fine-tuning compared to smaller encoder-only models in code search tasks?

\end{itemize}

These RQs systematically examine both the effectiveness and efficiency of decoder-only LLMs in code search. RQ1 establishes a baseline by assessing their zero-shot performance, while RQ2 investigates the impact of fine-tuning. RQ3 further explores the key factors driving performance improvements, considering fine-tuning methods, training data, model size, and input characteristics. Beyond accuracy, RQ4 examines computational cost, providing practical insights into model efficiency. Finally, RQ5 evaluates training efficiency, assessing how well these models adapt and generalize during fine-tuning. Together, these RQs provide a holistic evaluation of decoder-only LLMs for code search, balancing performance with resource efficiency.

\textbf{Results and Key Findings.}  
This study evaluates \revised{eleven} state-of-the-art (SOTA) decoder-only LLMs for code search tasks, conducting a comprehensive analysis of their performance across two fine-tuning methods, two types of datasets, and \revised{five} model sizes. Although decoder-only models initially underperform in zero-shot settings due to the mismatch between their code representations and the requirements of the code search task, fine-tuning significantly improves their performance, enabling them to better leverage their pre-trained code knowledge. Among the models evaluated, fine-tuned CodeGemma emerged as the top performer. On the CSN dataset, CodeGemma achieved a \revised{4.8\%} improvement in average MRR over the leading encoder-only model, UniXcoder. On the \cosqaplus dataset, where neither model was explicitly trained, CodeGemma demonstrated impressive gains by achieving a \revised{40.4\%} increase in MAP compared to UniXcoder. These findings highlight the effectiveness of decoder-only LLMs, especially after fine-tuning, in generalizing across unseen datasets, with their larger size and richer pretraining enhancing their ability to adapt to different code search scenarios.

Our analysis also indicates that fine-tuning on code-specific datasets, employing supervised contrastive learning, and \revised{mid-sized model} contribute to performance improvements. However, model architecture remains crucial, as larger models do not always guarantee better results. Decoder-only LLMs excel in long-code searches but struggle with ultra-short queries (fewer than 10 tokens) due to the curse of dimensionality and insufficient context. Although larger models lead to longer computational times, the costs are manageable, and they demonstrate superior training efficiency and generalization on limited data compared to smaller encoder-only models. In summary, this study highlights the significant potential of fine-tuned decoder-only LLMs for code search tasks, with strong performance and generalization across varying query lengths and datasets.

Our replication package, including code and results, is available online~\cite{ChenyxEugene_DecoderLLMs_CodeSearch_2025}. 
\revRtwo{Furthermore, to facilitate further research, all fine-tuned models evaluated in this study are publicly available in a dedicated Hugging Face Collection~\cite{SYSUSELab_HF_Collection_2025}. }
The key contributions of this work are:  

\begin{itemize}
    \item \textbf{First Systematic Study on Decoder-Only LLMs for Code Search}:  
    We systematically examine decoder-only LLMs in code search, comparing their zero-shot and fine-tuned performance.  

    \item \textbf{Comprehensive Benchmarking}:  
    We demonstrate that fine-tuned decoder-only LLMs, particularly CodeGemma, surpass encoder-only models like \unixcoder in code search, showcasing superior generalization.  

    \item \textbf{Optimization Strategies}:  
    We provide insights into model selection, fine-tuning techniques, training data quality, model size, query/code length impact, computational cost, and training efficiency.  
\end{itemize}

%% file: sections/background.tex
\section{Background}

\subsection{Code Search}
Early code search engines were based on keyword matching between queries and code, primarily relying on text similarity methods\cite{chatterjee2009sniff}. While functional, these approaches struggled to accurately capture the semantics of code due to significant differences between programming and natural languages. Recent advancements in deep learning have led to more sophisticated models capable of extracting high-level semantic representations, significantly improving code search performance\cite{gu2018deep}. Deep learning models utilize neural networks to uncover hidden features from data, which aids in generating semantic representations of both natural language and code\cite{watson2022systematic}.

Gu et al. \cite{gu2018deep} were pioneers in applying deep neural networks to embed code and queries into a shared vector space, measuring their similarity through vector distances. Since then, various model architectures have been applied to code search, including sequence models \cite{ye2020leveraging,shuai2020improving}, convolutional neural networks (CNN) \cite{li2020learning,ling2020adaptive}, tree neural networks \cite{wan2019multi}, graph models \cite{ling2021deep}, and Transformer-based models \cite{du2021single,zhu2020ocor}.

The development of pre-trained models on large-scale code datasets has also enhanced semantic understanding and search capabilities. For example, models like BERT \cite{devlin2018bert}, CodeBERT \cite{feng2020codebert}, GraphCodeBERT \cite{guo2020graphcodebert}, and UniXcoder \cite{guo2022unixcoder} have demonstrated impressive performance by training on bidirectional Transformers, where each token can attend to all others. Encoder-only architectures are typically favored for code search tasks due to their ability to handle code understanding better than decoder-only or encoder-decoder architectures \cite{csn}.
However, these models still face critical challenges: 
\begin{itemize} 
    \item \textbf{Poor Generalization}: Fine-tuned models often require task-specific adjustments, but fine-tuning datasets tend to be small, noisy, and narrowly sourced, which increases costs and can limit model performance on broader scenarios\cite{bui2021self,sun2022importance}. Additionally, most models in code search, such as UniXcoder\cite{guo2022unixcoder}, have fewer than 100 million parameters, which restricts their generalization and performance on unseen examples. 

    % \item \textbf{Limited Handling of Long Texts}: Many encoder-only models, such as CodeBERT\cite{feng2020codebert} and UniXcoder\cite{guo2022unixcoder}, are constrained by maximum input length limits. Queries and code snippets longer than these limits are truncated, potentially losing important context and diminishing the effectiveness of code search. 
\end{itemize}

Previous studies have shown that encoder-only models, such as UniXcoder, outperform smaller decoder-only models like CodeT5\cite{wang2021codet5} in code search tasks~\cite{icse2023ptmcomp}. However, the potential of decoder-only LLMs in this domain remains underexplored. This study aims to fill this gap by conducting the first systematic evaluation of decoder-only LLMs for code search. We compare their performance against traditional encoder-based models, offering insights into their strengths, limitations, and optimization strategies for improving code search performance.

\subsection{Decoder-only LLMs for Information Retrieval}
Recent research has applied decoder-only models to text embedding for information retrieval (IR), with notable improvements. For example, Ma et al. \cite{ma2024fine} fine-tuned LLaMA 2 \cite{touvron2023llama2} using S-BERT methods, and Wang et al. \cite{E5wang2023improving} created high-quality synthetic datasets for better IR. Springer et al. \cite{echoembeddingspringer2024repetition} proposed echo embeddings to address model robustness issues, while BehnamGhader et al. \cite{behnamghader2024llm2vec} used bidirectional attention and dual training sessions to improve performance.

However, these studies focus on general IR, with limited exploration in code search. No research has yet applied decoder-only LLMs like \deepseekllm, CodeLlama, or DeepSeekCoder to code search tasks. This gap highlights the need to explore the performance of decoder-only models in code search specifically.

To the best of our knowledge, this paper is the first to systematically explore decoder-only LLMs for code search. We demonstrate that models like CodeGemma outperform encoder-only models such as \unixcoder, offering superior generalization. Additionally, we provide insights into optimizing these models for code search, including model selection, fine-tuning, training data, and model size.

%% file: sections/study_design.tex
\section{Study Setup}
\label{sec:setup}

In this section, we present the benchmark, metrics, and the selection/configuration of the LLMs used in our study.
\input{sections/setup/benchmark}
\input{sections/setup/metrics}

\input{sections/setup/models}

%% file: sections/setup/benchmark.tex
\subsection{Benchmark}
\label{sec:benchmark}
To thoroughly assess the performance of decoder-only LLMs in code search tasks, we utilized the \csn (CSN)~\cite{csn} and \cosqaplus~\cite{gong2024cosqa+} datasets.

\begin{itemize}
    \item The CSN dataset is a comprehensive benchmark specifically designed for evaluating code search tasks~\cite{csn}, frequently used by previous work~\cite{shi2023cocosoda,codesurvey}. It encompasses a diverse range of programming languages, including Python, Java, JavaScript, Ruby, Go, and PHP. The dataset consists of millions of code snippets extracted from open-source repositories automatically, each paired with corresponding natural language documentation. This pairing facilitates the evaluation of code search models by measuring their ability to retrieve relevant code snippets in response to natural language queries.
    \item The \cosqaplus dataset is the latest benchmark for Python code search~\cite{gong2024cosqa+}. \cosqaplus is an enhanced version of the \cosqa~\cite{cosqa} dataset, designed to address common challenges in existing code search datasets. It improves upon \cosqa by pairing high-quality queries with multiple appropriate code snippets, blocks, and functions. These queries come from \cosqa, and the code snippets are sourced from the filtered StaQC~\cite{yao2018staqc} and \csn~\cite{csn} datasets. The candidate pairs are formed using multiple models and annotated automatically with the help of LLMs like Claude 3 Sonnet and GPT-4o, ensuring accurate matches.
    Using this dataset mitigates data leakage risk in two key ways: first, some of the generated responses do not originate from CSN, significantly reducing the potential for data leakage. Second, it differs from previous code search benchmarks in its task objectives by matching queries with multiple suitable code snippets. This divergence in dataset content and task goals further decreases the likelihood of data leakage.
\end{itemize}

We selected these two datasets due to their complementary strengths. CSN is widely used and covers multiple programming languages, facilitating alignment with previous research. \cosqaplus, released in 2024, provides a realistic and challenging benchmark for Python code search, featuring multiple correct answers per query. This combination allows us to evaluate our models comprehensively across different languages and complexity levels. Table~\ref{tab:dataset} provides more detailed information about these datasets, including the composition and number of datasets. 

\begin{table}[htbp]
  \centering
  \caption{Datasets Used for Evaluation}
    \vspace{-5pt} 
    \begin{tabular}{cp{3.2em}cccc}
    \toprule
    \multicolumn{1}{p{1.8em}}{\textbf{Dataset}} & \multicolumn{1}{p{2.8em}}{\textbf{Language}} & \multicolumn{1}{p{2.8em}}{\textbf{\#Train}} & \multicolumn{1}{p{2.8em}}{\textbf{\#Valid}} & \multicolumn{1}{p{2.8em}}{\textbf{\#Test}} & \multicolumn{1}{p{7.5em}}{\textbf{\#Candidate Code}} \\
    \midrule
    \multirow{6}[2]{*}{CSN} & Ruby  & 24,927 & 1,400 & 1,261 & 4,360 \\
          & JavaScript & 58,025 & 3,885 & 3,291 & 13,981 \\ 
          & Java  & 164,923 & 5,183 & 10,955 & 40,347 \\
          & Go    & 167,288 & 7,325 & 8,122 & 28,120 \\
          & PHP   & 241,241 & 12,982 & 14,014 & 52,660 \\
          & Python & 251,820 & 13,914 & 14,918 & 43,827 \\
    \midrule
    \cosqaplus & Python & 87,098    & 10,869  & 10,929 & 60,576 \\
    \bottomrule
    \end{tabular}%
  \label{tab:dataset}%
\end{table}%

%% file: sections/setup/metrics.tex
\subsection{Metrics}
\label{sec:metrics}

We used Mean Reciprocal Rank (MRR)~\cite{voorhees1999trec} and Mean Average Precision (MAP)~\cite{codesurvey} as our primary evaluation metrics. These metrics are widely adopted in information retrieval (IR)~\cite{hambarde2023information} and have been commonly used in previous code search research~\cite{codesurvey,shi2023cocosoda}. Both MRR and MAP are designed to be maximized, with their values ranging from 0 to 1, where higher values indicate better performance.

\parabf{MRR.} MRR measures the mean of the reciprocals of the rank positions of the first relevant code snippet for each query. It is calculated as shown in Equation~\ref{eq:mrr}, where $\text{Rank}_i$ denotes the rank position of the first relevant code snippet for the $i$-th query, and $N$ is the total number of queries. MRR is particularly useful in code search tasks because it emphasizes the rank of the first relevant result, reflecting the effectiveness of retrieving relevant code snippets early.

\begin{equation}
\label{eq:mrr}
\text{MRR} = \frac{1}{N} \sum_{i=1}^{N} \frac{1}{\text{rank}_i}
\end{equation}

\parabf{MAP.} MAP calculates the average precision score at each relevant item retrieved, averaged over multiple queries. It is computed as shown in Equation~\ref{eq:map}, where $N$ is the total number of queries, and $\text{AP}_i$ is the Average Precision for the $i$-th query. The Average Precision ($\text{AP}_i$) is calculated as shown in Equation~\ref{eq:ap}, where $Q_i$ is the total number of relevant code snippets for the $i$-th query, and $\text{rank}(i,j)$ denotes the rank position of the $j$-th relevant code snippet for the $i$-th query. MAP provides a comprehensive measure of search performance by considering the precision of relevant results across the entire result set. It is especially suited for evaluating datasets like \cosqaplus that involve multiple correct answers per query.

\begin{equation}
\label{eq:map}
\text{MAP} = \frac{1}{N} \sum_{i=1}^{N} \text{AP}_i
\end{equation}

\begin{equation}
\label{eq:ap}
\text{AP}_i = \frac{1}{Q_i} \sum_{j=1}^{Q_i} \frac{j}{\text{rank}(i,j)}
\end{equation}

%% file: sections/setup/models.tex
\subsection{Models}
\label{sec:models}

\input{tables/tab_llm}

To thoroughly evaluate the effectiveness of decoder-only LLMs in code search tasks, we selected \revised{11} state-of-the-art (SOTA) LLMs extensively examined in recent studies. We focused on open-source models released after 2023, excluding smaller models (with fewer than 1 billion parameters) due to their limited efficacy and larger models (with more than 7 billion parameters) due to computational resource constraints. Table \ref{tab:models} presents the details of the LLMs studied in our experiments, including model type, \revRtwo{base models, version, publication year, model sizes  and training data sizes.}  Our study encompasses a diverse range of decoder-only LLMs, varying across multiple dimensions such as (i) utilization of different base models, (ii) inclusion of various versions of series models (e.g., Llama2 and Llama3), and (iii) specialization in general-purpose versus code-specific training. To ensure readability throughout the paper, we refer to models by their common names (e.g., ``CodeGemma") after specifying the exact version used in the table.  

The selected LLMs are divided into two categories: general LLMs and code LLMs. General LLMs are trained for broad tasks, while code LLMs are specialized for software engineering, either trained from scratch on code corpora or fine-tuned with additional code data on top of general LLMs. We aim to investigate whether code LLMs outperform general LLMs in code search tasks, which require strong understanding of both natural language and code. For the general LLMs, we include their code-specific versions where available. Currently, a code-specific version of Llama3 is not available.

The details of the studied LLMs are as follows:

\begin{itemize}
    \item \textbf{Llama2}~\cite{touvron2023llama2}: Employs a decoder-only transformer \revRtwo{architecture} with a 32K token vocabulary. \revised{ While the larger 70B model in this family uses Grouped-Query Attention (GQA) for efficiency, the 7B and 13B versions used in our study utilize standard Multi-Head Attention. All versions feature Rotary Positional Embeddings.}
    % It features Grouped-Query Attention for efficient computation and Rotary Positional Embeddings for better positional understanding.
    \item \textbf{CodeLlama}~\cite{roziere2023code}: Developed by fine-tuning Llama2 with a higher sampling of code, featuring code infilling capabilities, support for large input contexts, and zero-shot instruction following for programming tasks.

    \item \textbf{Mistral}~\cite{jiang2023mistral}: Incorporates Grouped \revRtwo{Query} Attention for faster inference and Sliding Window Attention to handle longer sequences efficiently, enhancing contextual understanding and reducing memory usage.
    \item \textbf{CodeMistral}~\cite{uukuguy2023speechless}: Derived from Mistral, trained \revRtwo{on a curated, mixed dataset of approximately 201,981 samples from six sources (including Open-Platypus\cite{lee2023platypus}, OpenOrca\cite{Lian2023OpenOrca}, and WizardLM\cite{xu2024wizardlm}) specifically filtered for coding and reasoning tasks.}

    \item \textbf{DeepSeekLLM}~\cite{bi2024deepseek}: Follows the \revRtwo{Llama} model design with a Pre-Norm structure and RMSNorm function, using SwiGLU for feed-forward network activation.
    \item \textbf{DeepSeekCoder}~\cite{guo2024deepseek}: Initially pre-trained with a diverse dataset consisting of 87\% code, 10\% code-related language (e.g., GitHub Markdown and StackExchange), and 3\% non-code-related Chinese language, with further instruction fine-tuning.
    \item \textbf{Gemma}~\cite{team2024gemma}: A lightweight, text-to-text, decoder-only LLM from Google, designed for various text generation tasks and optimized for deployment in resource-limited environments.
    \item \textbf{CodeGemma}~\cite{team2024codegemma}: Further trained based on Gemma, using a combination of open-source math datasets and synthetically generated code to enhance logical reasoning and problem-solving skills.

    \item \textbf{Llama3}~\cite{grattafiori2024llama}: Utilizes a standard decoder-only transformer architecture with a 128K token vocabulary tokenizer and Grouped Query Attention (GQA) for improved efficiency and performance.
    \item \revised{\textbf{StarCoder2}~\cite{lozhkov2024starcoder}: Trained on over 4 trillion tokens of code from The Stack v2 (covering 17 programming languages), it utilizes Grouped Query Attention (GQA), a 16,384-token context window with a 4,096-token sliding window, and a Fill-in-the-Middle (FIM) training objective to enhance code understanding.}
    \item \revised{\textbf{\revRtwo{Qwen2.5-Coder}}~\cite{hui2024qwen2}: A series of multilingual code models from Alibaba's Qwen team, trained on a large-scale corpus of high-quality text and code, and supporting an extensive context window for better long-range dependency understanding.}

\end{itemize}

Additionally, we include two encoder-only models for comparison: CodeBERT \cite{feng2020codebert} and UniXcoder \cite{guo2022unixcoder}. CodeBERT, based on the RoBERTa architecture, is pre-trained on a large code corpus using Masked Language Modeling (MLM) and Replaced Token Detection (RTD). UniXcoder, on the other hand, is pre-trained with a combination of masked language modeling, unidirectional language modeling, denoising autoencoding, and contrastive learning on multi-modal data, including code summaries and abstract syntax trees.
We do not include smaller decoder-only models, such as CodeT5 \cite{wang2021codet5}, for comparison. Previous work \cite{icse2023ptmcomp} has shown that these models tend to perform worse than UniXcoder on code search tasks.

We obtained these models from their official repositories and used them for zero-shot inference or fine-tuning according to the provided guidelines. All evaluations were performed on an NVIDIA A800 80GB GPU.

%% file: tables/tab_llm.tex
\begin{table}[htbp]
  \centering
  \caption{Studied Models}
    \vspace{-5pt}
    \setlength{\tabcolsep}{4pt}
    \begin{tabular}{cccccc}
    \toprule
    {Type} & \multicolumn{1}{c}{{Model}} & \revised{Version} & \revRtwo{Publication} & {Size} & {Train. Data} \\
    \midrule
    \multicolumn{1}{c}{\multirow{5}[2]{*}{General LLM}} & Llama3 & Instruct & \revRtwo{2024}  & 8B    & 15000B \\
          & Mistral & Instruct & \revRtwo{2023}  & 7B    & - \\
          & DeepSeekLLM & Instruct & \revRtwo{2024}  & 7B    & - \\
          & Gemma & Instruct & \revRtwo{2024}  & 7B    & 500B \\
          & Llama2 & Instruct & \revRtwo{2023}  & 7B    & 2000B \\
    \midrule
    \multicolumn{1}{c}{\multirow{6}[2]{*}{Code LLM}} & Qwen2.5-Coder & Instruct & \revRtwo{2024}  & 7B    & 5500B \\
          & StarCoder2 & Base  & \revRtwo{2024}  & 7B    & 658B \\
          & CodeMistral & Instruct & \revRtwo{2023}  & 7B    & - \\
          & DeepSeekCoder & Instruct & \revRtwo{2024}  & 6.7B  & 2000B \\
          & CodeGemma & Instruct & \revRtwo{2024}  & 7B    & 500B \\
          & CodeLlama & Instruct & \revRtwo{2023}  & 7B    & 2000B \\
    \midrule
    \multirow{2}[2]{*}{Encoder-Only} & CodeBERT & N/A   & \revRtwo{2020}  & 250M  & 0.95B \\
          & UniXcoder & N/A   & \revRtwo{2022}  & 250M  & 156B \\
    \bottomrule
    \end{tabular}%
  \label{tab:models}%
\end{table}%

%% file: sections/rq1.tex
\section{RQ1: Zero-shot Performance} 
In this RQ, we investigate the performance of decoder-only LLMs on code search tasks in a zero-shot setting. This means applying the model directly to the code search task by obtaining embeddings of the query and the code in the code corpus without any task-specific fine-tuning. We compare the performance of eleven SOTA decoder-only LLMs and two SOTA encoder-only models under these conditions using the CSN and \cosqaplus datasets.

\input{sections/RQ1/rq1_design}

\input{sections/RQ1/rq1_results}

%% file: sections/RQ1/rq1_design.tex
\subsection{Design}
\label{sec:rq1-design}
\input{tables/zeroshot}

\begin{figure}[t]
    \centering
    \begin{minipage}[b]{1\linewidth}
        \centering
        \includegraphics[width=\textwidth]{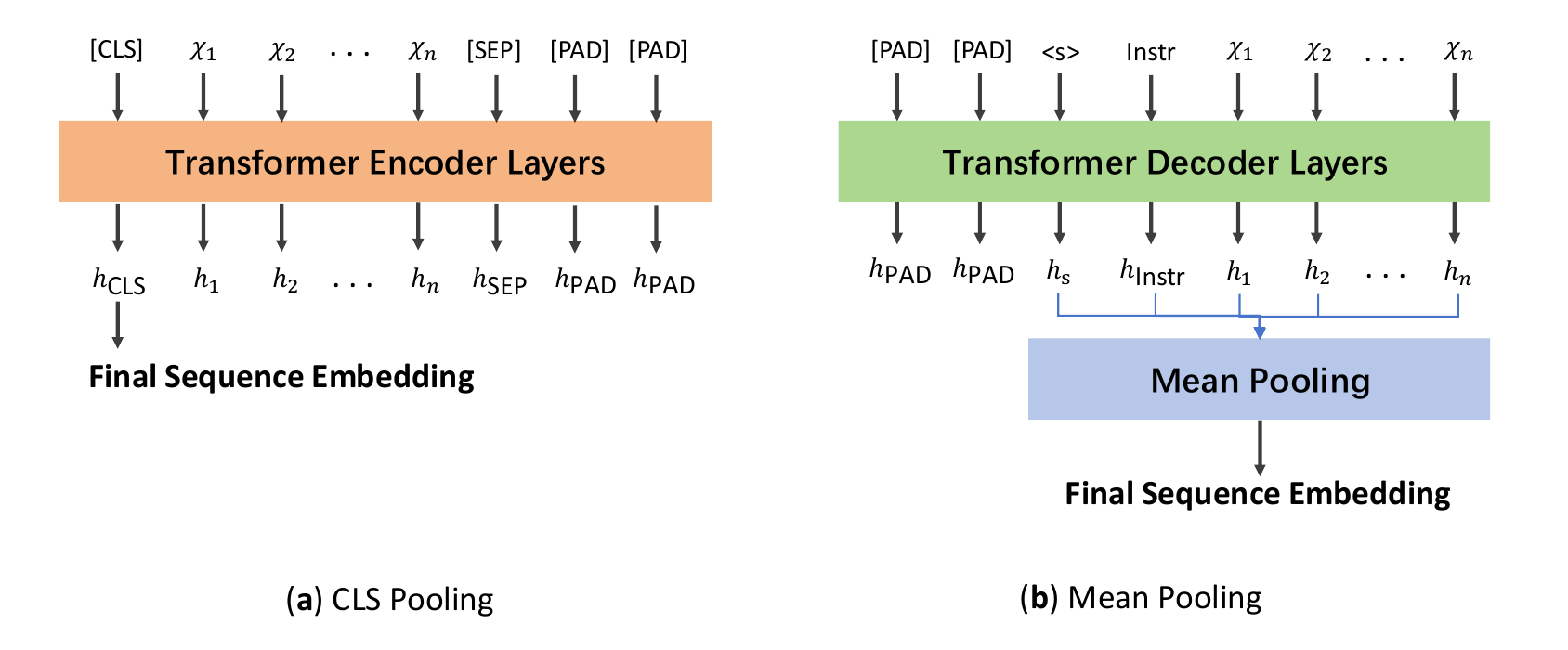}
        \vspace{-20pt} 
        \caption{Pooling Strategies in Encoder-Only and Decoder-Only Models}
        \label{fig:pooling_ways}
    \end{minipage}
\end{figure}

In this section, we outline the experimental design for evaluating the models using the CSN and \cosqaplus benchmarks. Our approach involves encoding both queries and code snippets into vector representations, followed by ranking the code snippets based on their cosine similarity to the query. We evaluate two categories of models: decoder-only LLMs (for both general-purpose and code tasks) and encoder-only models (i.e., UniXcoder and CodeBERT). The embeddings for queries and code snippets are generated based on the model category. \revRtwo{We set the consistent maximum input length to 512 tokens. This decision is based on a statistical analysis of the entire CodeSearchNet dataset (all six languages) using the UniXcoder tokenizer, which shows that this limit covers 99.46\% of queries and 92.79\% of code snippets.}

\parabf{Embeddings from Decoder-only LLMs.} For decoder-only LLMs, \revRtwo{ we employ mean pooling on the last hidden state to derive the sequence representation, a common practice for this architecture \cite{behnamghader2024llm2vec}. As illustrated in Figure \ref{fig:pooling_ways}(b), the process begins by concatenating an instruction with the input (query or code snippet) and applying left-padding. The entire sequence is then processed by the transformer layers. Finally, we compute the element-wise mean of all token embeddings from the last hidden state (excluding padding tokens) to produce the final sequence embedding. Across all experiments, we prepend each query with the following consistent instruction:`` Given a code search query, retrieve relevant passages that answer the query:". In contrast, code snippets are processed directly without any prefix. We adopt this approach of using a fixed instruction based on prior work, which indicates that model performance is not significantly sensitive to minor variations in \cite{echoembeddingspringer2024repetition}.}

\parabf{Embeddings from Encoder-only models.} For encoder-only models such as UniXcoder and CodeBERT, \revRtwo{we use the standard approach of selecting the [CLS] token’s embedding as the representation for the entire sequence \cite{ feng2020codebert, guo2022unixcoder}. Figure \ref{fig:pooling_ways}(a) depicts this process. The input sequence is formatted with a special [CLS] token at the beginning and right-padding at the end. After the model's transformer layers process the full sequence, the resulting output vector at the [CLS] position—which has aggregated information from the entire sequence—is taken as the final embedding.}

\parabf{Evaluation Methodology.}
Once the embeddings for the queries and code snippets are generated, we compute the cosine similarity between the query embedding and the embeddings of all code snippets in the dataset. The code snippets are then ranked based on their cosine similarity scores to the query.
We use MRR and MAP (see Section \ref{sec:metrics}) as our primary evaluation metrics, based on the ranking of ground truths for each query. MRR evaluates the effectiveness of retrieving relevant code snippets early in the ranking, while MAP provides a comprehensive measure of search performance by considering the precision of relevant results across the entire ranking.
For the CSN dataset, which contains a single ground truth per query, we calculated MRR. Since CSN includes multiple programming languages, we also evaluated performance across different programming languages.
For the \cosqaplus dataset, we calculated both MRR and MAP, as it includes multiple ground truths per query.
For both CSN and \cosqaplus, we only considered the top-1000 results when calculating MRR and MAP. Our evaluation approach follows the methodologies outlined in the official repositories of the CSN \cite{csn} and \cosqaplus \cite{gong2024cosqa+} benchmarks.

%% file: tables/zeroshot.tex
\begin{table*}[ht] 
  \centering
  \caption{Zero-Shot Performance on Code Search Benchmarks}
    \label{tab:zero-shot}%

    \begin{tabular}{|c|p{7.665em}|c|c|c|c|c|c|c|c|c|}
    \toprule
          & \multicolumn{1}{r|}{} & \multicolumn{7}{c|}{CSN(MRR)}                         & \multicolumn{2}{c|}{\revRtwo{\cosqaplus}} \\
    \midrule
    Category & \multicolumn{1}{c|}{Model} & Ruby  & \revRtwo{JavaScript} & Go    & Python & Java  & Php   & Avg.  & MAP   & MRR \\
    \midrule

    \multirow{2}[4]{*}{Encoder-Only} & CodeBERT & \revised{0.017 } & \revised{0.023 } & \revised{0.028 } & \revised{0.049 } & \revised{0.027 } & \revised{0.023 } & \revised{0.030 } & \revised{0.00051 } & \revised{0.00050 } \\
\cmidrule{2-11}          & UniXcoder & \revised{\textbf{0.619 }} & \revised{\textbf{0.634 }} & \revised{\textbf{0.696 }} & \revised{\textbf{0.744 }} & \revised{\textbf{0.731 }} & \revised{\textbf{0.682 }} & \revised{\textbf{0.706 }} & \revised{\textbf{0.17214 }} & \revised{\textbf{0.21065 }} \\
    \midrule
    \multicolumn{1}{|c|}{\multirow{5}[10]{*}{General LLMs}} & Llama3 & 0.134  & \textbf{0.214 } & \textbf{0.189 } & \textbf{0.352 } & \textbf{0.264 } & 0.167  & \textbf{0.239 } & \textbf{0.00803 } & \textbf{0.00928 } \\
\cmidrule{2-11}          & Mistral & 0.065  & 0.129  & 0.049  & 0.151  & 0.090  & 0.103  & 0.104  & 0.00357  & 0.00384  \\
\cmidrule{2-11}          & DeepSeekLLM & 0.075  & 0.143  & 0.048  & 0.154  & 0.138  & 0.091  & 0.114  & 0.00031  & 0.00032  \\
\cmidrule{2-11}          & Gemma & \textbf{0.163 } & 0.187  & 0.094  & 0.322  & 0.235  & \textbf{0.209 } & 0.222  & 0.00022  & 0.00023  \\
\cmidrule{2-11}          & Llama2 & 0.037  & 0.043  & 0.037  & 0.291  & 0.060  & 0.033  & 0.098  & 0.00040  & 0.00040  \\
    \midrule
    \multicolumn{1}{|c|}{\multirow{6}[12]{*}{Code LLMs}} & \multicolumn{1}{l|}{\revRtwo{Qwen2.5-Coder}} & \revised{\textbf{0.197 }} & \revised{\textbf{0.315 }} & \revised{\textbf{0.339 }} & \revised{\textbf{0.425 }} & \revised{\textbf{0.352 }} & \revised{\textbf{0.250 }} & \revised{\textbf{0.313 }} & \revised{\textbf{0.02003 }} & \revised{\textbf{0.01608 }} \\
\cmidrule{2-11}          & \multicolumn{1}{l|}{\revised{StarCoder2}} & \revised{0.099 } & \revised{0.168 } & \revised{0.117 } & \revised{0.202 } & \revised{0.173 } & \revised{0.120 } & \revised{0.154 } & \revised{0.00094 } & \revised{0.00094 } \\
\cmidrule{2-11}          & CodeMistral & 0.054  & 0.110  & 0.039  & 0.111  & 0.084  & 0.087  & 0.085  & 0.00040  & 0.00040  \\
\cmidrule{2-11}          & DeepSeekCoder & 0.107  & 0.140  & 0.105  & 0.191  & 0.180  & 0.106  & 0.147  & 0.00043  & 0.00043  \\
\cmidrule{2-11}          & CodeGemma & 0.155  & 0.125  & 0.086  & 0.204  & 0.088  & 0.076  & 0.114  & 0.00040  & 0.00041  \\
\cmidrule{2-11}          & CodeLlama & 0.081  & 0.153  & 0.068  & 0.117  & 0.131  & 0.100  & 0.110  & 0.00102  & 0.00102  \\

    \bottomrule
    \end{tabular}%
\end{table*}%

%% file: sections/RQ1/rq1_results.tex
\subsection{Results}
\label{sec:rq1-result}

Table \ref{tab:zero-shot} presents the zero-shot performance of various models on the CSN and \cosqaplus datasets. \revised{Bold indicates the best performance within each model category.}

\parabf{Best Model.} As shown in Table \ref{tab:zero-shot}, UniXcoder \revised{achieves} the highest zero-shot code search performance across multiple programming languages. It \revised{attains} an average MRR of \revised{0.706} on the CSN dataset, as expected, since UniXcoder was pre-trained on CSN and fine-tuned on NL-PL pairs. It also \revised{outperforms} other models on \cosqaplus, with MAP and MRR scores of \revised{0.17214} and \revised{0.21065}, respectively.

UniXcoder's strong performance is due to its effective code fragment embeddings, learned through multi-modal contrastive learning (MCL) and cross-modal generation (CMG) \cite{guo2022unixcoder}, making it well-suited for code search tasks. However, its zero-shot performance on \cosqaplus, which was not part of its pretraining, \revised{drops} to an MRR of \revised{0.21065}. This decline is likely due to differences in query style: CSN queries are derived from method docstrings, while \cosqaplus queries are based on web search queries. These variations in query structure likely contribute to the performance gap across the datasets.

\parabf{Encoder-only Models Vs. Decoder-only LLMs.} As shown in Table \ref{tab:zero-shot}, UniXcoder, an encoder-only model, \revised{outperforms} all decoder-only LLMs by a significant margin. \revRtwo{Qwen2.5-Coder}, the best-performing decoder-only LLM, \revised{achieves} the highest performance on both CSN and \cosqaplus. However, even \revRtwo{Qwen2.5-Coder}, which excels in general and code tasks, performs worse than UniXcoder on both datasets.

This advantage is expected, as UniXcoder is designed for code understanding. It uses pretraining tasks to learn semantic embeddings that represent code fragments effectively, allowing for direct similarity calculations. In contrast, decoder-only LLMs, trained for next-token prediction, are not optimized for code and produce embeddings that do not align well with the needs of code search, resulting in suboptimal performance. Springer et al. \cite{echoembeddingspringer2024repetition} note that causal attention in decoder-only LLMs hinders their ability to fully capture code context. Similarly, CodeBERT, despite being pretrained on the CSN dataset, struggles with code search tasks in zero-shot settings due to its embeddings not being fine-tuned for the task during pretraining~\cite{feng2020codebert}. On the other hand, because decoder-only LLMs benefit from large parameter sizes and extensive pretraining, their performance on code search in zero-shot settings \revised{is} better than that of CodeBERT.

\finding{In zero-shot settings, decoder-only LLMs generally underperform compared to encoder-only models like UniXcoder, due to a mismatch between their pretraining objectives and the specific needs of code search tasks.}

\parabf{General LLMs Vs. Code LLMs.} 
It is often assumed that code LLMs should outperform general LLMs in code search tasks because they are specifically designed to enhance the model of ability to understand code. However, the results \revised{show} that code LLMs do not always perform better than their general LLM counterparts. For instance, Llama2 and \deepseekllm benefited from additional training, which improved their performance on code search tasks. Conversely, the performance of Gemma and Mistral slightly decreased with additional training. This decline may be due to the fact that while additional training improved the models' understanding of code, it may have compromised their ability to effectively understand and process search queries, which are crucial for code search tasks. Successful code search requires a strong understanding of both code and natural language queries. \revised{It is worth noting that the top two performers, Qwen2.5-Coder and Llama3, are the latest models evaluated.  This suggests that continuous advancements in training data, scale, and architectural techniques are the primary drivers of state-of-the-art performance, rather than the model's categorical label alone.}

\finding{\revised{In zero-shot setting, A decoder LLM's specialization (General-Purpose vs. Code-Specific) is not a primary determinant of its code search performance. Instead, model recency, reflecting underlying technological advancements, shows a stronger correlation with state-of-the-art results.}}

%% file: sections/rq2/rq2.tex
\section{RQ2: Fine-tuning Improvement} 
\label{sec:rq2}
In this section, we explore whether fine-tuning can help decoder-only LLMs bridge the gap between their pre-trained representations and the specific requirements of code search tasks. Specifically, we investigate how fine-tuning on the CSN dataset improves performance compared to the zero-shot setting.

\input{tables/sim-finetune-result}

%% file: tables/sim-finetune-result.tex
\begin{table*}[htbp]
  \centering
  \caption{Performance of Fine-Tuned Models on Code Search Benchmarks}
  \vspace{-5pt}
    \begin{tabular}{|c|p{7.665em}|c|c|c|c|c|c|c|c|c|}
    \toprule

          & \multicolumn{1}{r|}{} & \multicolumn{7}{c|}{CSN(MRR)}                         & \multicolumn{2}{c|}{\revRtwo{\cosqaplus}} \\
    \midrule
    Category & \multicolumn{1}{c|}{Sup Model} & Ruby  & \revRtwo{JavaScript} & Go    & Python & Java  & Php   & Avg.  & MAP   & MRR \\
    \midrule
    \multirow{2}[4]{*}{Encoder-Only} & \multicolumn{1}{l|}{CodeBERT} & \revised{0.481 } & \revised{0.451 } & \revised{0.519 } & \revised{0.600 } & \revised{0.576 } & \revised{0.500 } & \revised{0.542 } & \revised{0.00481} & \revised{0.00546} \\
\cmidrule{2-11}          & \multicolumn{1}{l|}{UniXcoder} & \revised{0.573 } & \revised{0.567 } & \revised{0.697 } & \revised{0.684 } & \revised{0.703 } & \revised{0.636 } & \revised{0.667 } & \revised{0.04217} & \revised{0.04933} \\
    \midrule
    \multicolumn{1}{|c|}{\multirow{11}[22]{*}{Decoder-Only}} & CodeGemma & \textbf{0.662 } & 0.643 & 0.786  & 0.776  & 0.749  & 0.709 & \textbf{0.740 } & 0.24168  & 0.28357  \\
\cmidrule{2-11}          & Gemma & 0.642  & 0.614  & \textbf{0.799 } & 0.741  & 0.736  & 0.698  & 0.725  & \textbf{0.24592}  & \textbf{0.28577}  \\
\cmidrule{2-11}          & DeepSeekCoder & 0.608  & 0.639  & 0.708  & 0.752  & 0.741  & \textbf{0.723}  & 0.724  & 0.14260  & 0.17523  \\
\cmidrule{2-11}          & DeepSeekLLM & 0.610  & 0.608  & 0.744  & 0.719  & 0.742  & 0.720  & 0.719  & 0.01001  & 0.01132  \\
\cmidrule{2-11}          & \multicolumn{1}{l|}{\revised{Qwen2.5-Coder}} & \revised{0.645 } & \textbf{0.656} & \revised{0.772 } & \textbf{0.793} & \textbf{0.754} & \revised{0.688 } & \revised{0.718 } & \revised{0.19327} & \revised{0.23963} \\
\cmidrule{2-11}          & CodeLlama & 0.569  & 0.645  & 0.747  & 0.748  & 0.742  & 0.686  & 0.718  & 0.13011  & 0.16277  \\
\cmidrule{2-11}          & CodeMistral & 0.569  & 0.578  & 0.772  & 0.707  & 0.724  & 0.694  & 0.706  & 0.20876  & 0.24258  \\
\cmidrule{2-11}          & Llama3 & 0.550  & 0.579  & 0.733  & 0.718  & 0.709  & 0.681  & 0.694  & 0.19349  & 0.24008  \\
\cmidrule{2-11}          & Mistral & 0.562  & 0.571  & 0.732  & 0.684  & 0.722  & 0.686  & 0.692  & \revRtwo{0.17822} & \revRtwo{0.21212}   \\
\cmidrule{2-11}          & \revised{StarCoder2} & \revised{0.537 } & \revised{0.597 } & \revised{0.684 } & \revised{0.692 } & \revised{0.710 } & \revised{0.658 } & \revised{0.646 } & \revised{0.00611} & \revised{0.00712} \\
\cmidrule{2-11}          & Llama2 & 0.392  & 0.471  & 0.584  & 0.568  & 0.580  & 0.503  & 0.545  & 0.01592  & 0.01911  \\
    \bottomrule
    \end{tabular}%
  \label{tab:finetune_result}
\end{table*}%

%% file: sections/rq2/rq2-design.tex
\subsection{Design}
\label{sec:rq2:design}

\begin{figure}[t]
    \centering
    \begin{minipage}[b]{.99\linewidth}
        \centering
        \includegraphics[width=\textwidth]{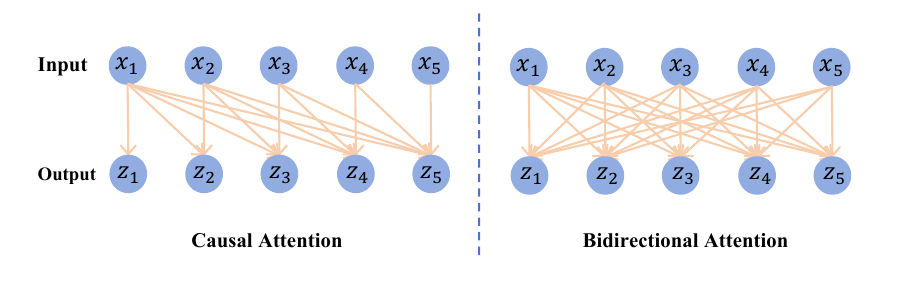}
        \vspace{-20pt} 
        \caption{Attention Mechanism}
        \label{fig:Attention}
    \end{minipage}
\end{figure}

\parabf{Training Approach.}
We fine-tuned the models using the CSN dataset and supervised contrastive learning (SupCon), following common practices \cite{guo2022unixcoder,codesurvey}. This method minimizes the distance between similar samples and maximizes the distance between different ones, ensuring relevant queries and code snippets are closely represented in the embedding space, while irrelevant ones are placed farther apart. Springer et al. \cite{echoembeddingspringer2024repetition} note that causal attention in decoder-only LLMs can hinder sentence-level understanding, leading to poor performance. To address this, we employed a bidirectional attention mechanism and incorporated Masked Next Token Prediction (MNTP) training prior to fine-tuning, as suggested by \cite{behnamghader2024llm2vec}.

As shown in Figure \ref{fig:Attention}, causal attention restricts the model to using information from the current token and its preceding context, limiting its ability to capture global semantic structures. In contrast, the bidirectional attention mechanism \cite{devlin2018bert} enables each token to access both preceding and subsequent contexts, improving the model's understanding of semantic relationships. MNTP training \cite{devlin2018bert,liu2019roberta} masks non-terminal tokens and predicts their content, helping the model learn complex syntactic structures and enhancing its compatibility with the bidirectional attention mechanism.

\parabf{Training Data.}
 Since CSN only provides positive samples, we constructed negative samples for each query by randomly sampling non-corresponding code snippets from the training dataset, as random negatives have shown robust and acceptable performance in previous work\cite{hasan2021codesc,feng2020codebert,guo2020graphcodebert}. We used this constructed dataset to conduct contrastive learning for all models.

\parabf{Training Setting.}
We trained all decoder-only LLMs, as listed in Table \ref{tab:models}, for 1000 steps with a batch size of 64, utilizing 2 NVIDIA A800 80GB GPUs for around 7.5 hours per model. For comparison, we applied the same data and approach to supervised contrastive learning for the CodeBERT and UniXcoder. To ensure fair comparison, we used the same random seed, training data, and hyperparameters for each model.

\revised{For the fine-tuning of our decoder-only LLMs,  we employed Parameter-Efficient Fine-Tuning (PEFT) using the LoRA technique with a rank ($\text{LoRA}_{r}$) of 16 to manage computational demands. The models were optimized using AdamW with a learning rate of 2e-4 and a linear warmup over the first 300 steps. To ensure consistency, we maintained a maximum input length of 512 tokens (as justified in Section IV) and trained each model for 1,000 steps.
This duration was primarily determined by our own empirical validation, as we observed that performance on a validation set ceased to improve significantly beyond this point. This protocol is also consistent with approaches in prior work \cite{behnamghader2024llm2vec}. To ensure computational efficiency, we utilized bfloat16 mixed-precision training, gradient checkpointing, and FlashAttention-2. The complete fine-tuning scripts, datasets, and all hyperparameter configurations are publicly available in our replication package~\cite{ChenyxEugene_DecoderLLMs_CodeSearch_2025}.}

\parabf{Evaluation Approach.}
The fine-tuned models \revRtwo{are} evaluated using the same metrics and test datasets as described in RQ1 (Section \ref{sec:rq1-design}) to assess their performance.

%% file: sections/rq2/rq2-result.tex
\subsection{\revised{Results}}
\label{sec:rq2:result}

\begin{figure*}[t]
    \centering
    \subfloat[\footnotesize \revRtwo{UniXcoder Before Fine-tuning} \\ \quad \label{fig:len_impact_of_query_length}]{
        \includegraphics[width=.23\textwidth]{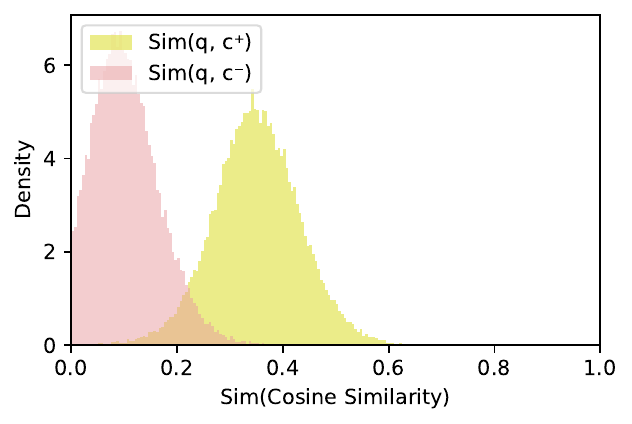}
    }
    \hspace{0.05em} 
    \subfloat[\footnotesize \revRtwo{UniXcoder After Fine-tuning }\label{fig:CodeGemma_Uni_rank}]{
        \includegraphics[width=.23\textwidth]{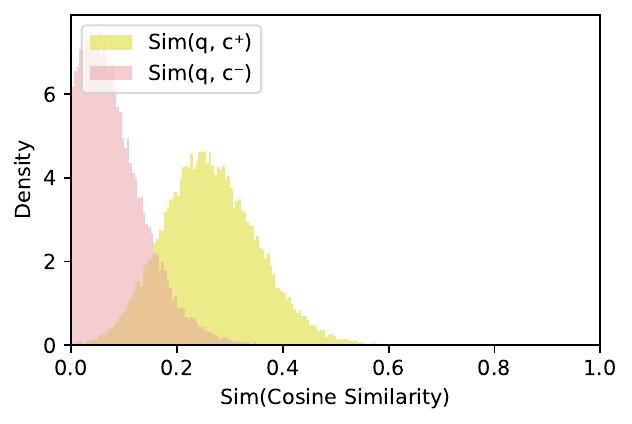}
    }
    \hspace{0.05em} 
    \subfloat[\footnotesize \revRtwo{CodeGemma Before Fine-tuning}  \\ \quad \label{fig:histogram_diff_av_top10}]{
        \includegraphics[width=.23\textwidth]{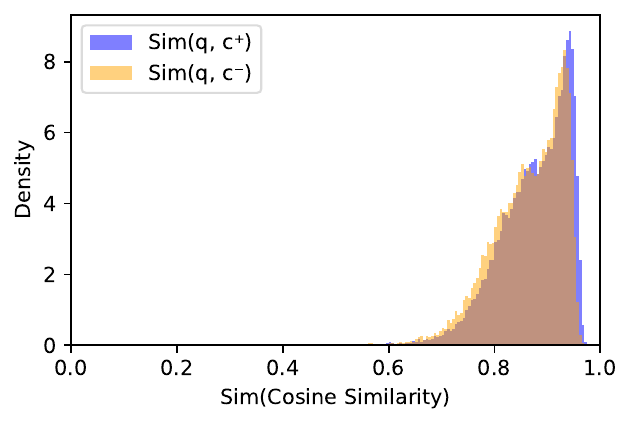}
    }
    \hspace{0.05em} 
    \subfloat[\footnotesize \revRtwo{CodeGemma After Fine-tuning} \label{fig:repeated_Q_sub}]{ 
        \includegraphics[width=.23\textwidth]{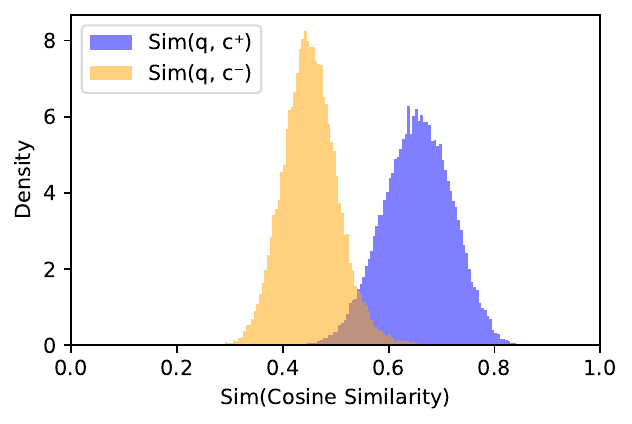}
    }
    \vspace{-12pt}
    \caption{\revRtwo{Similarity Histograms on \cosqaplus for UniXcoder and CodeGemma: Before and After Fine-tuning on CSN.}}
    \label{fig:all_four_histograms} 
\end{figure*}

Table~\ref{tab:finetune_result} presents the code search performance of the studied models after fine-tuning, with decoder-only LLMs sorted by average MRR across different programming languages on CSN.

\parabf{Impact of Fine-Tuning on Decoder-Only LLMs.}
Compared to the zero-shot setting, all decoder-only LLMs \revRtwo{show} substantial improvements after fine-tuning, with average MRR on CSN increasing by 453.1\% to 727.4\%. The improvement is even more pronounced on \cosqaplus, a dataset with a style significantly different from the CSN training data. This suggests that fine-tuning enables LLMs to fully leverage their pre-trained code understanding capabilities, aligning their vector representations with the specific needs of code search tasks. By fine-tuning, models adjust their existing understanding of code, narrowing the gap and making them better suited for similarity-based retrieval. Importantly, this improved capability is transferable, indicating that fine-tuned models can generalize to different code search tasks.

The ranking of models after fine-tuning differs significantly from the zero-shot results, highlighting that the benefits of fine-tuning are not uniform across models. In general, code-specific LLMs outperform general-purpose LLMs in code search tasks across all models we \revRtwo{evaluate}.
As shown in Table~\ref{tab:finetune_result}, CodeGemma and Gemma ranked first and second, respectively, on the CSN dataset, followed by DeepSeekCoder and DeepSeekLLM, while Mistral \revRtwo{rank} \revised{third} to last. This is noteworthy because Mistral is commonly used as a baseline in information retrieval research \cite{mteb_leaderboard}. It suggests that the code search domain may have unique characteristics that make models based on the Gemma architecture more effective.

\finding{Fine-tuning decoder-only LLMs enhances code search performance, narrowing the zero-shot gap and enabling models, especially code-specific ones, to better leverage pre-trained code understanding. The fine-tuned models also demonstrate good generalization, showing performance improvements on tasks with significant style differences, even without specific training.}

\parabf{Decoder-Only LLMs vs. Encoder-only Models.}
After fine-tuning, decoder-only LLMs \revRtwo{outperform} encoder-only models on both CSN and \cosqaplus. Notably, \unixcoder's performance declines post-fine-tuning, likely due to overfitting and the significant style difference between CSN and \cosqaplus. This suggests limitations in the generalization of small encoder-only models like \unixcoder.

In contrast, decoder-only LLMs \revRtwo{demonstrate} notable improvements, with some achieving SOTA performance. For example, \revRtwo{the} fine-tuned DeepSeekCoder, Gemma, and CodeGemma \revRtwo{show} improvements of \revised{2.5\%, 2.6\%, and 4.8\%} in average MRR on CSN compared to zero-shot UniXcoder. Additionally, CodeGemma \revRtwo{exhibits} a \revised{40.4\%} improvement in MAP and a \revised{34.6\%} improvement in MRR on \cosqaplus over zero-shot UniXcoder. This highlights that the larger size and pre-training of decoder-only LLMs enable them to generalize better to unseen datasets like \cosqaplus, even without specific training.

Among the models, CodeGemma \revRtwo{demonstrates} exceptional post-fine-tuning performance, surpassing both Gemma and other models across most metrics on CSN, \revRtwo{and while its performance on \cosqaplus ranks second, it also remains close to that of Gemma.} This suggests that additional code pre-training enhanced its code understanding. Although this initially impacted its search task performance, fine-tuning significantly \revRtwo{boosts} its capabilities.

\finding{Fine-tuned decoder-only LLMs, particularly CodeGemma, outperform encoder-only models, showcasing superior generalization and achieving SOTA performance on both CSN and \revRtwo{top-tier performance on \cosqaplus.}}

\subsection{\revised{Qualitative Analysis}}

A puzzling result from our experiments is that fine-tuning UniXcoder on the CSN dataset, while improving its in-domain performance, leads to a significant performance drop on the \cosqaplus benchmark. This is particularly intriguing as the same fine-tuning process significantly \revised{boosts} the performance of decoder-only models like CodeGemma. To understand the root cause of these divergent outcomes, we conduct a comparative \revRtwo{quantitative analysis to measure the change in discriminative power for both models.}

\revRtwo{\textbf{Analysis Design.} To measure the models' ability to distinguish relevant from irrelevant code, we design an analysis on the \cosqaplus benchmark. For each query in the benchmark, we define its corresponding code snippet as the positive sample (q, c$^+$) and a code snippet randomly sampled from the rest of the benchmark as a negative sample (q, c$^-$). We then compute the cosine similarity for a large population of positive and negative pairs for both UniXcoder and CodeGemma, before and after fine-tuning.}

\revRtwo{\parabf{Analysis of Similarity Distributions (Figure \ref{fig:all_four_histograms}).} The results of this analysis are visualized in Figure \ref{fig:all_four_histograms} , which plots the density histograms of positive (Sim(q, c$^+$)) and negative (Sim(q, c$^-$)) samples similarities. To quantify the separation, we also compute the mean difference between positive and negative similarity scores.}

\revRtwo{\parabf{UniXcoder (Figures \ref{fig:all_four_histograms}a, \ref{fig:all_four_histograms}b).} Before fine-tuning (Figures \ref{fig:all_four_histograms}a), UniXcoder exhibits strong discriminative power. The positive and negative distributions are visibly separated, with the positive pairs (yellow) centered at a higher similarity than the negative pairs (red). At this stage, the model has a mean similarity difference of 0.251. However, after fine-tuning on CSN (Figures \ref{fig:all_four_histograms}b), the model's ability to handle this out-of-domain (OOD) task appears to degrade. The two distributions move noticeably closer, with a larger area of overlap. This is confirmed quantitatively, as the mean similarity difference drops to 0.206. This objectively demonstrates that fine-tuning on CSN may have diminished UniXcoder's ability to discriminate between relevant and irrelevant code. This result suggests that the model has likely overfitted to the in-domain CSN data, which could cause it to lose its discriminative power for the more diverse, OOD \cosqaplus queries.}

\revRtwo{\parabf{CodeGemma (Figures \ref{fig:all_four_histograms}c, \ref{fig:all_four_histograms}d).} In contrast, CodeGemma shows a contrasting behavior. Before fine-tuning (Figures \ref{fig:all_four_histograms}c), the base CodeGemma model has almost no discriminative power; the positive (orange) and negative (blue) distributions are nearly identical, with a negligible mean difference of 0.012. This helps explain its poor initial performance. However, after fine-tuning (Figures \ref{fig:all_four_histograms}d), the model learns to organize the space effectively. The distribution for positive pairs (blue) shifts clearly to the right (higher similarity), while the negative distribution (orange) remains centered at a lower similarity. This creates a distinct separation, and the mean difference increases substantially from 0.012 to 0.201.}

\finding{\revRtwo{The impact of in-domain fine-tuning on OOD generalization appears to be architecture-dependent. For smaller encoder models like UniXcoder, this process may reduce OOD discriminative power, suggesting overfitting. In contrast, for decoder-only LLMs like CodeGemma, it substantially enhances this power, potentially improving generalization.}}

%% file: sections/RQ3/RQ3.tex
\section{RQ3: Improvement Analysis} 
\label{sec:rq3}

In this RQ, we explore the factors driving performance gains from fine-tuning decoder-only LLMs in code search tasks, focusing on four key aspects: training method, training data, model size, and query and code length.

\input{sections/RQ3/training-approach}

\input{sections/RQ3/training-data}

\input{tables/mults_vs_single}
\input{sections/RQ3/Single-Language-Fine-Tuning}

\input{sections/RQ3/model-size}

\input{figures/rq3-figures}

\input{sections/RQ3/advantage-limitation}

%% file: sections/RQ3/training-approach.tex
\subsection{Training Method}
\label{sec:train-Mothod}

\begin{table}[t]
  \centering
  \caption{Results of Different Fine-Tuning Methods}
  \vspace{-5pt} 
    \begin{tabular}{|c|p{4.18em}|c|c|c|}
    \toprule
    Model & \multicolumn{1}{c|}{Fine-tuning} & CSN(Avg.) & CoS.(MAP) & CoS.(MRR) \\
    \midrule
    \multirow{3}[6]{*}{CodeGemma} & \multicolumn{1}{c|}{Zero-Shot} & 0.114  & 0.00040  & 0.00041  \\
\cmidrule{2-5}          & SimCSE & 0.077  & 0.00021  & 0.00023  \\
\cmidrule{2-5}          & SupCon   & \textbf{0.740 }\revised{*} & \textbf{0.24168}\revised{*}  & \textbf{0.28357 }\revised{*} \\
    \midrule
    \multirow{3}[6]{*}{Llama3} & Zero-Shot & 0.239  & 0.00803  & 0.00928  \\
\cmidrule{2-5}          & SimCSE & 0.391  & 0.02894  & 0.03759  \\
\cmidrule{2-5}          & SupCon   & 0.694\revised{*}  & 0.19349\revised{*}  & 0.24008\revised{*}  \\
    \midrule
    \multirow{3}[6]{*}{Mistral} & Zero-Shot & 0.104  & 0.00357  & 0.00384  \\
\cmidrule{2-5}          & SimCSE & 0.257  & 0.02972  & 0.03679  \\
\cmidrule{2-5}          & SupCon   & 0.692\revised{*}  & \revRtwo{0.17822}\revised{*} &  \revRtwo{0.21212}\revised{*}  \\
    \bottomrule
    \end{tabular}%
  \label{tab:Training Approach}%
\end{table}%

\begin{figure*}[t]
    \centering
    \subfloat[\scriptsize CodeGemma]{
    \includegraphics[width=0.31\textwidth]{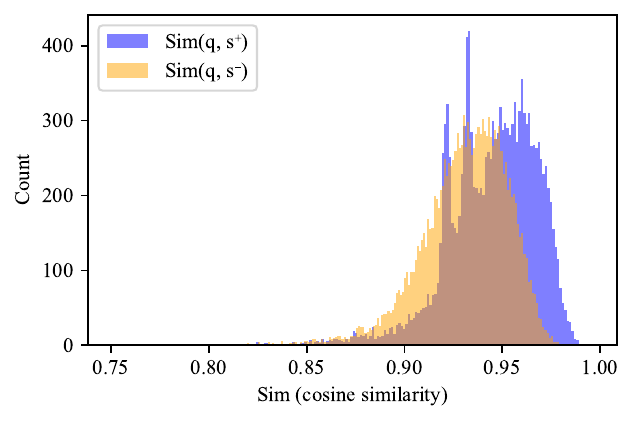}}
    % \quad
    \hspace{0.3em}  
    \subfloat[\scriptsize SimCSE CodeGemma]{
    \includegraphics[width=0.31\textwidth]{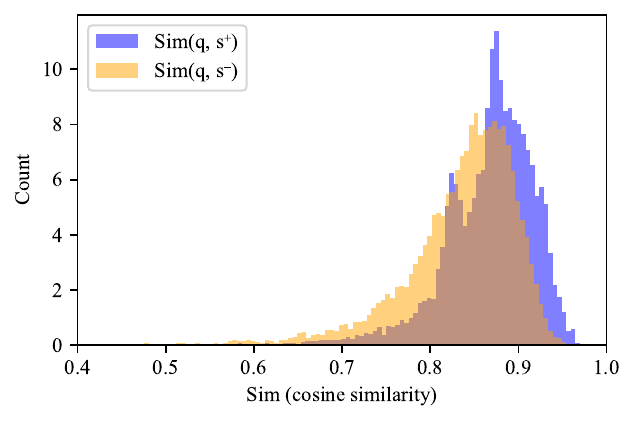}}
    % \quad
    \hspace{0.3em}  
    \subfloat[\scriptsize SupCon CodeGemma]{
    \includegraphics[width=0.31\textwidth]{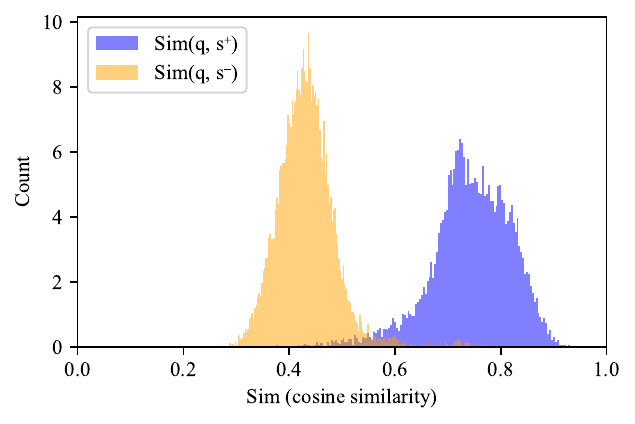}}
    % \quad
    \vspace{0pt} 
    \centering
    \caption{Similarity Histograms of Different Fine-Tuning Methods}
    \label{fig:CodeGemma_similarity_histogram}
\end{figure*}

To understand the impact of training methods, we compare supervised contrastive learning, as used in RQ2 (refer to Section \ref{sec:rq2:design}), with unsupervised contrastive learning, specifically SimCSE \cite{gao2021simcse}, and analyze their effectiveness.

\parabf{Comparison Method.} SimCSE, an unsupervised contrastive learning approach, uses dropout masks to generate independent samples of the same input sentence. It aims to maximize the similarity between these samples while minimizing similarity with other sentences. We \revRtwo{follow} the approach in \cite{behnamghader2024llm2vec} and \revRtwo{use} the Wikitext-103 dataset for training. As Wikipedia data is included in the pre-training of all models, it does not impart new knowledge. For a fair comparison, models \revRtwo{are} adapted to a bidirectional attention mechanism and \revRtwo{undergo} MNTP training, similar to the setup in Section \ref{sec:rq2:design}.

\parabf{Selected Models.} We \revRtwo{select} CodeGemma, Llama3, and Mistral for comparison. CodeGemma \revRtwo{is} chosen for its top test results, Llama3 for being the latest model, and Mistral for being the most widely used decoder-only language model in information retrieval applications \cite{mteb_leaderboard}.

\parabf{\revised{Statistical Analysis.}}
\revised{To verify the statistical significance of our findings, we compare the performance of different approaches using the \textbf{Wilcoxon signed-rank test}, a non-parametric paired test. We report a result as statistically significant if the p-value is less than our chosen significance level of \textbf{$\alpha = 0.05$}. To account for multiple comparisons across different models and datasets, we apply the \textbf{Benjamini-Hochberg (BH) correction} to the p-values. In Table \ref{tab:Training Approach}, results marked with an asterisk (*) meet this significance criterion of $p < 0.05$.}

\parabf{Results and Analysis.} 
Table \ref{tab:Training Approach} shows the performance comparison of CodeGemma, Llama3, and Mistral on CSN and \cosqaplus. The results indicate that supervised contrastive learning significantly outperforms unsupervised contrastive learning (SimCSE) across all three models. However, SimCSE also improves the performance of models comparing to the zero-shot setting in some cases.

To assess the impact of the training method on model performance, we \revRtwo{select} the top-performing models and \revRtwo{focus} on the Go language, which exhibits the highest MRR performance for CodeGemma in CSN. We \revRtwo{analyze} the cosine similarity of vectors generated by model embeddings across three methods: zero-shot, SimCSE, and supervised contrastive learning. Using all CSN test queries in Go, we \revRtwo{treat} matched code as positive samples and randomly \revRtwo{select} other code as negative samples. Details of this analysis for the Go language are presented in Fig. \ref{fig:CodeGemma_similarity_histogram}. 
In each subplot, the blue histograms represent the cosine similarity between the query and positive samples, while the orange histograms represent the similarity between the query and negative samples. Higher x-axis values correspond to greater similarity, and the height of the bars indicates the frequency of occurrences at each similarity level. 
\revised{Figure \ref{fig:CodeGemma_similarity_histogram} visualizes CodeGemma's ability to differentiate between relevant (positive) and irrelevant (negative) code snippets under our different experimental settings.}

The results in Fig.\ref{fig:CodeGemma_similarity_histogram}  reveal that while unsupervised contrastive learning improves performance, it does not significantly enhance the distinction between positive and negative samples. Supervised contrastive learning, however, greatly improves the correlation between queries and positive samples while reducing the correlation with negative samples. Note that we \revRtwo{perform} the same analysis for other programming languages in CSN, and the trends \revRtwo{are} consistent. However, due to space limitations, we do not present these results here.

\finding{Supervised contrastive learning outperforms unsupervised contrastive learning in fine-tuning decoder-only LLMs for code search by better distinguishing positive and negative samples, leading to improved performance.}

%% file: sections/RQ3/training-data.tex
\subsection{Training Data}
\label{Training_Data}

\input{tables/analysis-training-data}

In this section, we analyze the impact of different training data on model performance. 
We used the E5 dataset and the CSN dataset. The E5 dataset \cite{E5wang2023improving} is a universal query dataset with approximately 1.5 million samples covering 93 languages, making it the most diverse dataset, focusing on general information retrieval tasks rather than code search. 

For similar reasons to RQ3.a in Section \ref{sec:rq3}, we select CodeGemma, Llama3, and Mistral for supervised contrastive learning. For a fair comparison, all the training settings are the same for both datasets, as detailed in Section \ref{sec:rq2:design}. \revised{ The statistical significance of the results is evaluated using the same protocol described in Section \ref{sec:train-Mothod} (Wilcoxon signed-rank test with BH correction).}

\revised{Table \ref{tab:train data} confirms that fine-tuning on the specialized CSN dataset generally leads to a performance advantage over the general-purpose E5 dataset. This trend is most pronounced for CodeGemma, where the improvement is statistically significant across all metrics. However, our statistical analysis reveals that the magnitude of this advantage is highly dependent on the base model and evaluation task. For a powerful model like Llama3, the benefit of specialized data is not statistically significant on the in-domain CSN benchmark, though it remains significant for the more complex \cosqaplus task. This suggests that while specialized data is beneficial, its impact can be modulated by the extensive knowledge already present in state-of-the-art pretrained models.}

\finding{\revised{While specialized code search datasets generally provide a performance boost, its statistical significance and magnitude are highly dependent on the base model. The benefit is universally significant for some models (e.g., CodeGemma), but can be marginal for more powerful models (e.g., Llama3) on certain tasks, offering a crucial trade-off for practitioners.}}

%% file: tables/analysis-training-data.tex
\begin{table}[t]
  \centering
  \caption{Results of Different Fine-Tuning Datasets}
  \vspace{-5pt} 
    \begin{tabular}{|c|p{2.835em}|c|c|c|}
    \toprule
    \multicolumn{1}{|p{6.39em}|}{SupCon Model} & Dataset & CSN(Avg.) & CoS.(MAP) & CoS.(MRR) \\
    \midrule
    \multirow{2}[4]{*}{CodeGemma} & E5    & 0.689  & 0.04309  & 0.05124  \\
\cmidrule{2-5}          & CSN   & \textbf{0.740\revised{*} } & \textbf{0.24168}\revised{*}  & \textbf{0.28357\revised{*} } \\
    \midrule
    \multirow{2}[4]{*}{Llama3} & E5    & 0.666  & 0.16996  & 0.21939  \\
\cmidrule{2-5}          & CSN   & 0.694  & 0.19349\revised{*}  & 0.24008\revised{*}  \\
    \midrule
    \multirow{2}[4]{*}{Mistral} & E5    & 0.667  & 0.16663  & 0.20709  \\
\cmidrule{2-5}          & CSN   & 0.692\revised{*}  & \revRtwo{0.17822} & \revRtwo{0.21212}  \\
    \bottomrule
    \end{tabular}%
  \label{tab:train data}%
\end{table}%

%% file: tables/mults_vs_single.tex
\begin{table*}[htbp]
  \centering
  \caption{\revised{CodeGemma: Multi- vs. Single-language Tuning for Code Search}}
  \vspace{-5pt}

    \begin{tabular}{|c|c|c|c|c|c|c|c|c|c|c|}
    \toprule
    
    \multicolumn{2}{|c|}{\revised{}} & \multicolumn{7}{c|}{\revised{CSN(MRR)}} & \multicolumn{2}{c|}{\revised{\cosqaplus}} \\
    \midrule
    \revised{Model} & \revised{Training Language} & \revised{Ruby} & \revised{\revRtwo{JavaScript}} & \revised{Go} & \revised{Python} & \revised{Java} & \revised{Php} & \revised{Avg.} & \revised{MAP} & \revised{MRR} \\
    \midrule
    \multicolumn{1}{|c|}{\multirow{7}[14]{*}{\revised{\revRtwo{CodeGemma}}}} & \revised{Ruby} & \revised{0.656 } & \revised{0.623 } & \revised{0.690 } & \revised{0.760 } & \revised{0.728 } & \revised{0.712 } & \revised{0.717 } & \revised{0.14846 } & \revised{0.17173 } \\
\cmidrule{2-11}          & \revised{\revRtwo{JavaScript}} & \revised{0.607 } & \revised{\textbf{0.691 }} & \revised{0.701 } & \revised{0.768 } & \revised{\textbf{0.761 }} & \revised{\textbf{0.755 }} & \revised{\textbf{0.744 }} & \revised{0.12916 } & \revised{0.15050 } \\
\cmidrule{2-11}          & \revised{Go} & \revised{0.575 } & \revised{0.593 } & \revised{0.769 } & \revised{0.712 } & \revised{0.720 } & \revised{0.674 } & \revised{0.701 } & \revised{0.01942 } & \revised{0.02168 } \\
\cmidrule{2-11}          & \revised{Python} & \revised{0.361 } & \revised{0.455 } & \revised{0.499 } & \revised{0.552 } & \revised{0.490 } & \revised{0.460 } & \revised{0.491 } & \revised{0.00197 } & \revised{0.00217 } \\
\cmidrule{2-11}          & \revised{Java} & \revised{0.526 } & \revised{0.564 } & \revised{0.626 } & \revised{0.656 } & \revised{0.701 } & \revised{0.693 } & \revised{0.665 } & \revised{0.10881 } & \revised{0.12580 } \\
\cmidrule{2-11}          & \revised{Php} & \revised{0.482 } & \revised{0.521 } & \revised{0.509 } & \revised{0.619 } & \revised{0.572 } & \revised{0.516 } & \revised{0.552 } & \revised{0.04173 } & \revised{0.04753 } \\
\cmidrule{2-11}          & \revised{Multi-language} & \revised{\textbf{0.662 }} & \revised{0.643 } & \revised{\textbf{0.786 }} & \revised{\textbf{0.776 }} & \revised{0.749 } & \revised{0.709 } & \revised{0.740 } & \revised{\textbf{0.24168 }} & \revised{\textbf{0.28357 }} \\

    \bottomrule
    \end{tabular}%
  \label{tab:mults_vs_single}%
\end{table*}%

%% file: sections/RQ3/Single-Language-Fine-Tuning.tex
\subsection{Single-Language Fine-Tuning}
\label{sec:rq3_single_lang}

\revised{Our previous analysis \revRtwo{establishes} that fine-tuning on a specialized dataset yields superior performance. This finding, however, motivates a deeper investigation, as the specialized CSN corpus is itself multilingual. It remains unclear how the specific programming language used for fine-tuning influences a model's ability to learn generalizable code representations. To explore this, we \revRtwo{fine-tune our top-performing model, CodeGemma,} on six single-language subsets derived from CSN. All other fine-tuning configurations, as detailed in Section \ref{sec:rq2:design}, \revRtwo{are} kept identical to ensure a fair comparison.}

\revised{The results, presented in Table \ref{tab:mults_vs_single}, reveal that multilingual training provides superior generalization, challenging the straightforward assumption that monolingual fine-tuning is always the optimal path for language-specific specialization.}

\revised{On one hand, the multi-language model demonstrates remarkable generalization and the power of positive knowledge transfer. It achieves a competitive average MRR across all languages on the CSN benchmark and exhibits a commanding, cliff-like performance advantage on the more comprehensive \cosqaplus benchmark. This suggests that exposure to a diverse set of languages enables the model to learn more abstract and robust semantic representations of code. Furthermore, the multi-language model surprisingly outperforms single-language models on their own respective test sets for languages like Go, Python, and \revRtwo{Ruby}, indicating that the collective knowledge transferred from a diverse linguistic pool can provide a more powerful learning signal than specializing on a single language's dataset alone.}

\revised{On the other hand, monolingual fine-tuning exhibits significant instability and unpredictable results. Contrary to expectations, a model fine-tuned on a single language does not consistently achieve the best performance on its corresponding test set. In our experiments, the model fine-tuned solely on \revRtwo{JavaScript} not only \revRtwo{achieves} the top score on the \revRtwo{JavaScript} benchmark but also \revRtwo{matches} the best-performing specialized models on the Java and PHP test sets, all while securing the highest overall average MRR on CSN. This indicates that fine-tuning on \revRtwo{JavaScript} promotes a deep understanding that extends beyond \revRtwo{JavaScript} itself to the entire CSN corpus. In contrast, the model trained exclusively on Ruby \revRtwo{achieves} the best performance of all single-language models on the \cosqaplus benchmark.}

\revised{Furthermore, we observe that the choice of a single language for fine-tuning is a critical factor, as the effectiveness of each language as a data source varies dramatically. For instance, the model fine-tuned solely on \revRtwo{JavaScript} \revRtwo{achieves} the highest average MRR of any model, indicating its utility as a strong source for learning general code features. In stark contrast, the model trained only on Python \revRtwo{performs} poorly across all benchmarks, suggesting that Python's prevalence in the initial training corpora may have contributed to diminishing returns during fine-tuning, leading to less competitive performance compared to other languages.}

\finding{\revised{Multilingual fine-tuning provides superior generalization and more reliable performance. Monolingual fine-tuning is an unstable strategy for specialization. Contrary to intuition, specializing \revRtwo{in} a single language does not guarantee the best performance for that language.}}

\input{tables/discarding_language}

\revised{While the previous section \revRtwo{confirms} the benefits of multilingual training, it is crucial to understand how individual languages contribute to this overall success. This leads to a critical question regarding data composition: What is the performance impact of reducing or removing a language from the training mix? We address this through a new experiment that systematically discards data for selected languages from the CSN fine-tuning set.}

\revised{We \revRtwo{choose} Java and Ruby for this study to represent two distinct cases: Java, a high-resource language, and Ruby, a comparatively lower-resource language. Apart from the modifications to the training corpus, all other fine-tuning parameters \revRtwo{are} kept identical to those detailed in Section \ref{sec:rq2:design}.}

\revised{The results are presented in Table \ref{tab:discarding_language}, where \texttt{Discard Language} indicates the language being removed, and \texttt{Discard Ratio} specifies the proportion of its data that is discarded (0 for complete retention, 1 for complete removal). For clarity and to directly illustrate the impact of data removal, we have simplified the CSN evaluation metrics. \texttt{On Discarded Lang.} shows the model's performance on the test set of the language being discarded (e.g., the Java test set when Java is the discard language). \texttt{Avg. on Other Langs.} represents the average performance on the remaining languages' test sets. This presentation intuitively links the act of discarding a language's data to its direct and indirect performance consequences.}

\revised{
As shown in the table, the baseline configuration (a discard ratio of 0) consistently yields the best performance on the \cosqaplus benchmark and strong results on CSN. This reaffirms our earlier conclusion that the complete multilingual dataset is optimal for model generalization. However, a more nuanced, non-linear pattern emerges as we incrementally discard data. We observe that the poorest performance often occurs at intermediate discard ratios, rather than at the extremes. For instance, when discarding Java, the weakest CSN result appears at a ratio of 0.8, while the \cosqaplus performance dips lowest
at
\revRtwo{a ratio of 1.0}. Similarly, for Ruby, both benchmarks show the worst performance at a 0.5 ratio.}

\revised{We hypothesize that this may be due to a ``critical mass" effect. When the training data for a particular language falls below a certain threshold but is not absent, it may be insufficient to form a coherent representation and is instead treated by the model as statistical ``noise." This noise can disrupt the learning process for other languages more significantly than the language's complete absence (a ratio of 1). Furthermore, the On Discarded Lang. column reveals a crucial distinction tied to pre-training knowledge. For a lower-resource language like Ruby, performance degrades substantially as its specialized fine-tuning data is removed. In contrast, for a high-resource language like Java, the model maintains strong performance even when its fine-tuning data is completely discarded, likely because the model has already been exposed to a vast amount of Java code during its initial pre-training phase.}

\finding{\revised{The complete multilingual dataset generally provides stronger generalization. A small amount of data from one language appears to risk being misinterpreted by the model, potentially acting as statistical noise that disrupts the overall learning process.}}

%% file: tables/discarding_language.tex
\begin{table*}[htbp]
  \centering
  \caption{\revised{Performance of Fine-Tuned CodeGemma Models by Discarding Language-Specific Data}}
  \vspace{-5pt}
    \begin{tabular}{|c|c|c|c|c|c|c|}
    \toprule
    \revised{} & \multicolumn{2}{c|}{\revised{SupCon}} & \multicolumn{2}{c|}{\revised{CSN(MRR)}} & \multicolumn{2}{c|}{\revised{\cosqaplus}} \\
    \midrule
    \revised{Model} & \revised{Discard Language} & \revised{Discard Ratio} & \revised{On Discarded Lang.} & \revised{Avg. on Other Langs.} & \revised{MAP} & \revised{MRR} \\
    \midrule
    \multirow{10}[20]{*}{\revised{CodeGemma}} & \multirow{5}[10]{*}{\revised{Java}} & \revised{0} & \revised{0.749 } & \revised{0.715 } & \revised{\textbf{0.24168 }} & \revised{\textbf{0.28357 }} \\
\cmidrule{3-7}          &       & \revised{0.2} & \revised{0.724 } & \revised{0.661 } & \revised{0.13335 } & \revised{0.15156 } \\
\cmidrule{3-7}          &       & \revised{0.5} & \revised{0.689 } & \revised{0.651 } & \revised{0.14437 } & \revised{0.16190 } \\
\cmidrule{3-7}          &       & \revised{0.8} & \revised{0.674 } & \revised{0.611 } & \revised{0.16725 } & \revised{0.19062 } \\
\cmidrule{3-7}          &       & \revised{1} & \revised{\textbf{0.759 }} & \revised{\textbf{0.720 }} & \revised{0.10054 } & \revised{0.11544 } \\
\cmidrule{2-7}          & \multirow{5}[10]{*}{\revised{Ruby}} & \revised{0} & \revised{\textbf{0.662 }} & \revised{0.733 } & \revised{\textbf{0.24168 }} & \revised{\textbf{0.28357 }} \\
\cmidrule{3-7}          &       & \revised{0.2} & \revised{0.635 } & \revised{\textbf{0.736 }} & \revised{0.18360 } & \revised{0.20964 } \\
\cmidrule{3-7}          &       & \revised{0.5} & \revised{0.556 } & \revised{0.646 } & \revised{0.13991 } & \revised{0.16219 } \\
\cmidrule{3-7}          &       & \revised{0.8} & \revised{0.569 } & \revised{0.702 } & \revised{0.17533 } & \revised{0.20299 } \\
\cmidrule{3-7}          &       & \revised{1} & \revised{0.567 } & \revised{0.713 } & \revised{0.20265 } & \revised{0.23731 } \\

    \bottomrule
    \end{tabular}%
  \label{tab:discarding_language}%
\end{table*}%

%% file: sections/RQ3/model-size.tex
\subsection{Model Size}
\label{sec:rq3_Model_Size}

\begin{table}[t]
  \centering
  \caption{Results of Different Model Sizes}
  \vspace{-5pt}
\begin{tabular}{|c|p{3.72em}|c|c|c|}
\toprule
\multicolumn{1}{|p{6.39em}|}{SupCon Model} & Size(B) & \multicolumn{1}{p{4.165em}|}{CSN(Avg.)} & CoS.(MAP) & CoS.(MRR) \\
\midrule
\multicolumn{1}{|c|}{\multirow{3}[6]{*}{Llama2}} & 1.3B  & 0.458\revised{*}   & \textbf{0.03397 } & \textbf{0.04211 } \\
\cmidrule{2-5}  & 7B& \textbf{0.545 } & 0.01592\revised{*}  & 0.01911\revised{*}  \\
\cmidrule{2-5}  & 13B   & \revised{0.507 }\revised{*} & \revised{0.01516 }\revised{*} & \revised{0.01820 }\revised{*} \\
\midrule
\multirow{5}[10]{*}{\revRtwo{Qwen2.5-Coder}} & \revised{0.5B} & \revised{0.712 }\revised{*}  & \revised{0.19920 }\revised{*}  & \revised{0.23051 }\revised{*}  \\
\cmidrule{2-5}  & \revised{1.5B} & \revised{\textbf{0.741 }} & \revised{\textbf{0.24541 }} & \revised{\textbf{0.29551 }} \\
\cmidrule{2-5}  & \revised{3B} & \revised{0.719 }\revised{*}  & \revised{0.20342 }\revised{*}  & \revised{0.24796 }\revised{*}  \\
\cmidrule{2-5}  & \revised{7B} & \revised{0.718 }\revised{*}  & \revised{0.19327 }\revised{*}  & \revised{0.23963 }\revised{*}  \\
\cmidrule{2-5}  & \revised{14B} & \revised{0.695 }\revised{*}  & \revised{0.14351 }\revised{*}  & \revised{0.17694 }\revised{*}  \\
\bottomrule
\end{tabular}%
  \label{tab:model_size_results}%
\end{table}%

\revised{To investigate the impact of model size on code search performance, we \revRtwo{conduct} experiments across two different model families. We \revRtwo{fine-tune} models from the \textbf{Llama2 family} (using the 1.3B Sheared-LLaMA variant, 7B, and 13B) and the \textbf{Qwen2.5-Coder} family~\cite{Qwen2.5-Coder-Collection} (0.5B, 1.5B, 3B, 7B, and 14B) on the CSN dataset using our supervised contrastive learning setup.
An identical training setup \revRtwo{is} used for all model sizes to ensure a fair comparison (see Section \ref{sec:rq2:design} for details).}

\parabf{\revised{Statistical Analysis.}} \revised{As this analysis involves comparing more than two related groups (i.e., the different model sizes within a family), we \revRtwo{employ} a two-stage statistical procedure. First, we \revRtwo{use} the \parabf{ Friedman test} to determine if any statistically significant differences exist among the groups. If the result \revRtwo{is} significant ($p < 0.05$), we then \revRtwo{conduct} a post-hoc Wilcoxon test with Holm-Bonferroni correction for all pairwise comparisons to identify which specific pairs differed significantly. The significance level ($\alpha$) \revRtwo{is} set to 0.05. In our results tables, an asterisk (*) indicates that a model's performance is statistically significantly different from that of the best-performing model within the same comparison group ($p<0.05$).}

\revised{The results, summarized in Table \ref{tab:model_size_results}, reveal a nuanced relationship between model size and performance, challenging the simple assumption that ``bigger is always better".}

\revised{Our key observations are as follows:
\begin{itemize}
\item \textbf{Performance is not monotonic with size.} Our results challenge the simple ``bigger is better" scaling hypothesis within the context of code search. As shown in Table \ref{tab:model_size_results}, the Qwen2.5-Coder family exhibits a clear performance peak with the 1.5B variant, which statistically significantly outperforms both its smaller and larger counterparts.

\item \textbf{An optimal size ``sweet spot" appears to exist for tested architectures.} The Llama2 family shows a mixed trend\revRtwo{; specifically, its performance on the in-domain CSN dataset peaks at the 7B model, while the smallest 1.3B model achieves the best results on the out-of-domain \cosqaplus benchmark.}
This, combined with the clear peak for Qwen2.5-Coder, suggests that for the model architectures we investigated, an optimal model size may exist for this task, beyond which performance can degrade.

\item \textbf{Model architecture remains a critical factor.} The Qwen2.5-Coder architecture demonstrates a clear performance advantage, as its 1.5B model achieves substantially higher scores than any model from the Llama2 family on all metrics we tested.

\end{itemize}}

\finding{\revised{The relationship between model size and code search performance appears to be non-monotonic for the models under investigation. This highlights that, for specialized tasks like code search, simply scaling up parameters may not be the most effective strategy for improving performance.}}

%% file: figures/rq3-figures.tex
\begin{figure*}[t]
    \centering
    \centering
    \begin{minipage}[b]{.22\linewidth}
        \centering
        \includegraphics[width=\textwidth]{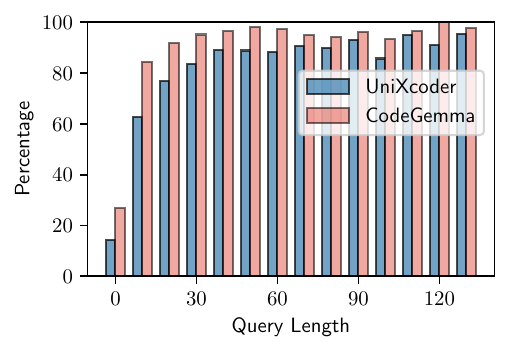}
        \vspace{-20pt} 
        \caption{Histogram of Query \quad \quad Length \\ \quad}
        \label{fig:len_impact_of_query_length}
    \end{minipage}\quad
    \begin{minipage}[b]{.22\linewidth}
        \centering
        \includegraphics[width=\textwidth]{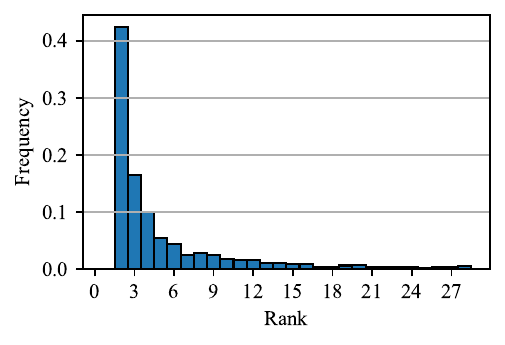}
        \vspace{-20pt} 
        \caption{\revised{UniXcoder Ranks on CodeGemma's Exact Matches }}
        \label{fig:CodeGemma_Uni_rank}
    \end{minipage}\quad
    \begin{minipage}[b]{.22\linewidth}
        \centering
        \includegraphics[width=\textwidth]{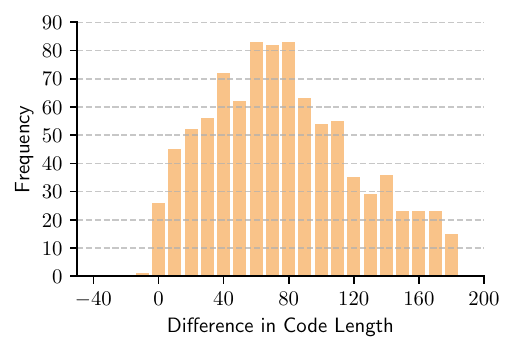}
        \vspace{-20pt} 
        \caption{Histogram of Difference in Code Length \\ \quad}
        \label{fig:histogram_diff_av_top10}
    \end{minipage}\quad
    \begin{minipage}[b]{.22\linewidth}
        \centering
        \includegraphics[width=\textwidth]{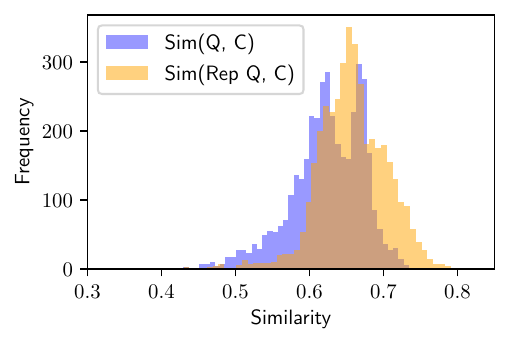}
        \vspace{-20pt} 
        \caption{Similarity Histogram of Different Queries}
        \label{fig:repeated_Q}
    \end{minipage}
\end{figure*}

%% file: sections/RQ3/advantage-limitation.tex
\subsection{Query and Code Length}
\label{sec:rq3_Query_and_Code_Length}

We \revRtwo{explore} whether different model architectures influence a model's preference between queries and code. We \revRtwo{analyze} this from two perspectives: the length of the query and code, and cases where the models match exactly on queries, i.e., top-1 ground truth hitting. We \revRtwo{examine} \unixcoder and CodeGemma after CSN fine-tuning, referring to them as \unixcoder and CodeGemma.

Some previous works have suggested that the past poor performance of decoder-only models in query results is due to the high-dimensional vector outputs causing curse of dimensionality which means that the data becomes sparse, making it difficult for the model to learn meaningful patterns \cite{mann2020language}. This \revRtwo{leads} us to investigate the token length of queries and code encoded by the models. We \revRtwo{analyze} the query accuracy of \unixcoder and CodeGemma tokens in different length intervals. As shown in Fig \ref{fig:len_impact_of_query_length}, both models perform poorly on ultra-short queries (queries with fewer than 10 tokens), but CodeGemma performs better on longer queries. This suggests that the larger size of the decoder-only model may lead to a better understanding of long queries.

\begin{figure}[t]
    \centering
    \includegraphics[width=0.5\textwidth]{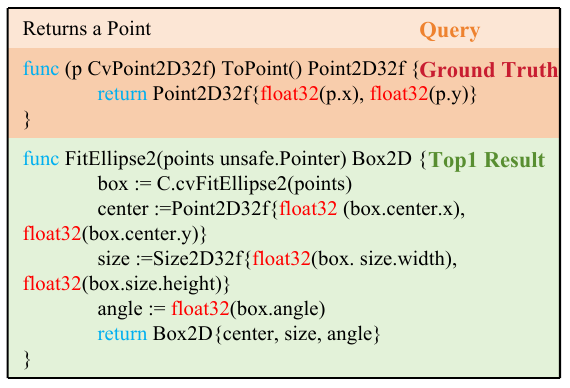}
    \vspace{-20pt} 
    \caption{Ultra-Short Queries Without Context}
    \label{fig:exmple_code}

\end{figure}

We \revRtwo{analyze} the difference in perfectly matching queries between \unixcoder and CodeGemma, finding that 36\% of the results matched perfectly only in CodeGemma.

However, this alone does not indicate bias. We further \revRtwo{analyze} the results in \unixcoder for these queries that only match perfectly in CodeGemma. Fig. \ref{fig:CodeGemma_Uni_rank} shows that the matching code often ranks second in the corresponding query in \unixcoder. In fact, 82\% of queries that match perfectly only in CodeGemma rank within the top 10 for the corresponding query in \unixcoder. This indicates that while the performance of decoder-only models is better than that of \revRtwo{Encoder-only} models, there is no inherent ``preference" in the embedding vector. The improvement is more about the model's enhanced understanding of code and search.

We \revRtwo{analyze} the search results based on CodeGemma and \revRtwo{find} that ultra-short queries (less than 10 tokens) often match longer code snippets. Fig. \ref{fig:histogram_diff_av_top10} shows that for ultra-short Go language queries with code tokens less than 50, 93\% of the top 10 matched codes are significantly longer, averaging 111 more tokens. This suggests that short query-short code combinations are weak spots for LLMs. Our case analysis \revRtwo{identifies} two main reasons for poor performance in ultra-short queries: 1. Sparse embedding vectors due to few tokens, leading to curse of dimensionality; 2. Low-quality matching results due to lack of context and unclear semantics.

To test the first reason, we \revRtwo{conduct} a small experiment by repeating queries to double their token length without changing semantics. We \revRtwo{test} ultra-short Go language queries on CSN using Llama3. Fig. \ref{fig:repeated_Q} shows that this method effectively \revRtwo{improves} the cosine similarity between queries and code embeddings. The results of the repeated queries \revRtwo{shift} to the right, indicating that the high-dimensional vectors of the decoder-only model can lead to curse of dimensionality, which can be mitigated by increasing the query length without altering the meaning.

Regarding the second reason, our case study of Go data with fewer than 10 query tokens and poor results \revRtwo{find} that such ultra-short queries often lack context and clear semantics. Fig. \ref{fig:exmple_code} shows an example of such low-quality queries with its top-1 incorrect result. We \revRtwo{observe} that ultra-short queries tend to favor longer code snippets due to the higher likelihood of keyword matches. For instance, the top result for this ultra-short query is a code snippet of 80 tokens, while the corresponding code token length is 37.

This issue might be \revRtwo{mitigated} through query reformulation techniques. By expanding or rephrasing ultra-short queries, we can provide additional context and improve the clarity of the query semantics. This could potentially mitigate the sparse embedding vectors problem and enhance the quality of matching results. For instance, adding contextually relevant terms or rephrasing vague terms can help align the query better with the intended information, leading to more accurate and relevant search results.

\finding{Fine-tuned decoder-only LLMs excel in long-code searches but struggle with ultra-short queries (fewer than 10 tokens) due to (1) curse of dimensionality from high-dimensional embeddings and (2) lack of context and semantic clarity.}

%% file: sections/RQ4/RQ4.tex
\section{RQ4: Computational Time Analysis} 
\label{sec:rq4}

The trade-off between model size and computational time is a critical consideration in the design and deployment of code intelligence systems. As shown in RQ2, while decoder-only LLMs can outperform small-scale encoder-only models in terms of performance and generalization on code search tasks after fine-tuning, one potential drawback is the increased computational time due to their larger number of parameters. This could affect their practical usability.

In this research question, we focus on investigating the computational time required for decoder-only LLMs in comparison to small-scale encoder-only models when applied to code search tasks.

\subsection{Design}

To investigate the trade-off between model size and computational time, we compare two models with significantly different parameter sizes: UniXcoder (125M parameters) and CodeGemma (7B parameters).

The typical workflow for applying these models includes the following steps:
\begin{itemize}
    \item Fine-tuning the model on the specific task (offline).
    \item Using the model to convert all code snippets in a codebase into code embeddings (offline).
    \item Indexing the code embeddings and creating an index using tools such as FAISS\footnote{https://github.com/facebookresearch/faiss} or Milvus\footnote{https://milvus.io/} (offline).
    \item Converting the query into the query embedding using the model (online).
    \item Searching the top-k most relevant code snippets using the query embedding from the index (online).
\end{itemize}

For the purpose of this study, we focus on comparing the per-query embedding time, per-code embedding time, and fine-tuning time. Other steps in the workflow, such as indexing and search, are not directly influenced by the models themselves and are therefore excluded from our analysis. Both models are trained and evaluated using the CSN benchmark, consistent with the settings described in RQ2 (Section~\ref{sec:rq2:design}). The per-query and per-code embedding times are calculated by averaging the total embedding time across all queries and code snippets, respectively.

\subsection{Results}

\begin{table}[t]
  \centering
  \caption{Computational Cost Comparison of UniXcoder and CodeGemma}
  \vspace{-5pt}
  \resizebox{\columnwidth}{!}{
  \begin{tabular}{|l|c|c|c|c|}
    \toprule
    Metric  & UniXcoder & CodeGemma & Ratio & Phase \\
    \midrule
    Size & 125M & 7B & 56x&- \\
        \midrule
    Fine-Tuning (min)  & 2.5 & 246.6 & 98.64× & Offline\\
    \midrule
    Per-Query Embedding (ms)  & 3.6 & 14.7 & 4.1× & Online\\
    \midrule
    Per-Code Embedding (ms)  & 4.2 & 32.6 & 7.8×  & Offline\\
    \bottomrule
  \end{tabular}%
  }
  \label{tab:runtime}%
\end{table}

Table \ref{tab:runtime} compares the computational costs of UniXcoder and CodeGemma across three key stages: fine-tuning, per-query embedding, and per-code embedding. The results highlight notable differences in computational demands, underscoring the trade-offs between model size and efficiency.

\parabf{Fine-tuning Time.}
In the fine-tuning phase, the time required for CodeGemma is significantly higher than for UniXcoder, with a 98.64× increase. However, this increase is an acceptable trade-off, as fine-tuning is an offline process and does not need to be repeated frequently. The performance and generalization improvements that come with using a larger decoder-only LLM like CodeGemma justify the additional fine-tuning time. Moreover, once fine-tuned, these models can be reused across various domains without the need for retraining, making the long fine-tuning time less of an issue in practical scenarios.

\parabf{Per-query Embedding Time.}
For per-query embedding, CodeGemma takes 14.7 ms, which is 4.1× longer than UniXcoder’s 3.6 ms. While this increase in embedding time is evident, it is important to note that the per-query embedding phase is relatively lightweight compared to the other stages, and thus the increased computational cost remains manageable. Additionally, this increase in time is still justifiable when considering the superior performance that CodeGemma offers, especially in terms of generalization on code search tasks. As the query embedding occurs in the online phase, where embeddings are computed in real-time as queries are issued, the additional time cost is minimal relative to the benefits.

\parabf{Per-code Embedding Time.}
The per-code embedding time for CodeGemma (32.6 ms) is substantially longer than UniXcoder (4.2 ms), reflecting a 7.8× increase. \revRtwo{
This difference is a direct consequence of CodeGemma's significantly larger and more complex architecture, featuring more layers and a wider hidden state than UniXcoder. Although both models process inputs truncated to the same 512-token length, CodeGemma's deeper architecture requires more computation per token. However, this cost is acceptable for two main reasons. First, code embeddings are computed offline and typically only once for a given codebase. Second, techniques such as data parallelism can be employed to significantly reduce the overall wall-clock time required for embedding a large codebase with sufficient hardware. Thus, while the per-code embedding time is longer for CodeGemma, it is a reasonable cost when considering its one-time calculation and its potential for significant performance improvements.}

In summary, while CodeGemma's larger model size results in longer computational times than UniXcoder, these costs remain manageable. The increased fine-tuning time is a one-time expense justified by improved performance and generalization, while the per-query and per-code embedding times, though longer, are reasonable given that embeddings are computed only once. 
\revRtwo{The longer embedding times are acceptable given that code indexing is an offline process that can be expedited with parallelization techniques.}

\finding{Decoder-only LLMs' larger model size results in longer computational times, but the increased costs are manageable.}

%% file: sections/rq6.tex
\section{RQ5: Training Efficiency}
\input{figures/RQ6_fig}

A key assumption behind the use of decoder-only LLMs is that their pretraining on large corpora and with numerous parameters has endowed them with a robust understanding of code. Consequently, these models may require fewer training samples to transfer their pre-existing knowledge to new tasks. To test this assumption, we compare the training efficiency and performance of decoder-only LLMs against smaller encoder-only models during fine-tuning on code search tasks.

\subsection{Design}
As in previous RQs, we compare CodeGemma (a decoder-only LLM) with UniXcoder, a leading encoder-only model. Both models are trained on the CSN training dataset, following the same training settings as described in RQ2 (see Section~\ref{sec:rq2}). We save the model checkpoints at regular intervals (every 200 training steps) and evaluate performance on the CSN test dataset as well as the \cosqaplus dataset to assess generalization. This comparison helps us understand how quickly each model can leverage available training data and how performance evolves as training progresses.

\subsection{Results}
Figure~\ref{fig:finetune_step_CSN} and Figure~\ref{fig:finetune_step_COSQA_MAP} illustrate the performance evaluations of the models at every 200 training steps (from 0 to 1000) on the CSN and \cosqaplus datasets, respectively. Our experiments reveal that UniXcoder shows a slight decline in performance between 200 and 800 training steps on the CSN dataset (see Figure~\ref{fig:finetune_step_CSN}), which suggests potential overfitting. This indicates that the model may rely too heavily on specific patterns from the limited fine-tuning samples, hindering its ability to generalize effectively. A similar trend is observed on the \cosqaplus dataset (Figure~\ref{fig:finetune_step_COSQA_MAP}), where UniXcoder’s performance plateaus after fine-tuning, with minimal improvement in adapting to new query styles and diverse tasks.

In contrast, CodeGemma shows significant improvements even after just 200 training steps (Figure~\ref{fig:finetune_step_CSN} and \ref{fig:finetune_step_COSQA_MAP}). As training progresses, CodeGemma continues to demonstrate consistent performance gains on both CSN and \cosqaplus, indicating that larger models like CodeGemma can better generalize from limited data, likely due to their larger pre-trained models and greater representational capacity. These results suggest that decoder-only LLMs, particularly larger ones like CodeGemma, can leverage contrastive learning with fewer samples to activate the knowledge learned during pre-training. This capability allows them to quickly improve performance on new tasks with minimal fine-tuning, demonstrating greater training efficiency.

These findings emphasize the advantages of decoder-only LLMs in terms of training efficiency and generalization. Thanks to their pretraining on large corpora and their larger parameter sizes, these models require fewer task-specific examples and adapt more quickly to new tasks. This makes them highly efficient for fine-tuning, particularly in code search applications, where data is often limited.

\finding{Decoder-only LLMs exhibit superior training efficiency and generalization on limited data compared to smaller encoder-only models.}

%% file: figures/RQ6_fig.tex
\begin{figure}[t]
    \centering
    \begin{minipage}[b]{.47\linewidth}
        \centering
        \includegraphics[width=\textwidth]{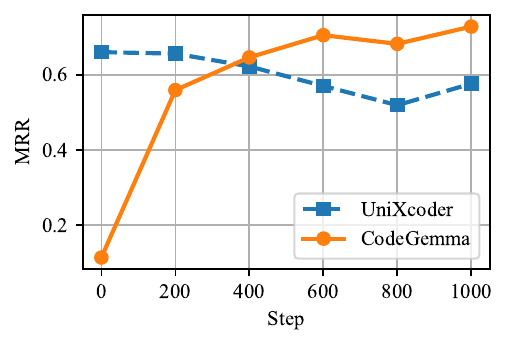}
        \vspace{-20pt} 
        \caption{Fine-tuning Performance of CodeGemma and UniXcoder on CSN}
        \label{fig:finetune_step_CSN}
    \end{minipage}\quad
    \begin{minipage}[b]{.47\linewidth}
        \centering
        \includegraphics[width=\textwidth]{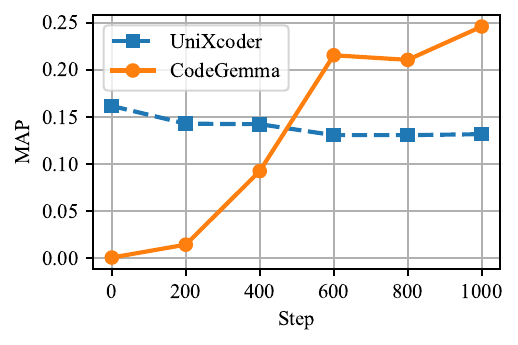}
        \vspace{-20pt} 
        \caption{Fine-tuning Performance of CodeGemma and UniXcoder on \cosqaplus}
        \label{fig:finetune_step_COSQA_MAP}
    \end{minipage}
\end{figure}

%% file: sections/RelatedWork.tex
\section{Related work} \label{sec:related_work}

\subsection{Broader Applications of Decoder-Only LLMs} 
\revRtwo{Recent research has begun to adapt decoder-only LLMs for various representation learning tasks, often achieving significant improvements. For example, in sequential recommendation, several frameworks like PAD \cite{wang2025pre} and LLM2Rec \cite{ he2025llm2rec} have been proposed to enhance traditional recommenders with LLM embeddings, aiming to solve cold-start problems or blend semantic and collaborative data. In multilingual embedding, LUSIFER presents a zero-shot approach to adapt English-centric LLM embedders for multilingual tasks \cite{man2025lusifer}.}

\revRtwo{While these studies expand the application of decoder-only LLMs in related embedding-centric domains, their specific application and fine-tuning behavior for the code search task remain largely unexplored.}

\subsection{Empirical Studies in Information Retrieval} 

\revRtwo{Concurrently, a growing body of empirical research has begun to dissect the performance, methodologies, and new challenges of LLMs in text retrieval. A recent comprehensive survey systematically overviews this paradigm shift from encoders to decoders for embedding generation, covering data, architecture, and pooling strategies \cite{tao2024llms}. Meanwhile, analytical studies have begun to explore the underlying model behaviors. For example, a recent analytical study \cite{feng2025learning} investigates how enabling bidirectional attention impacts the semantic representations of decoder-only LLMs for embedding tasks. Other studies analyze different facets, such as the safety risks of instruction-following retrievers \cite{behnamghader2025exploiting}.}

\revRtwo{To the best of our knowledge, there is still a significant gap in the literature regarding the systematic, empirical application of decoder-only LLMs to code search tasks. Our work addresses this gap by providing insights into model selection, data, and sizing for this specific domain.}

%% file: sections/threat.tex
\section{Threats \revRtwo{to} Validity}

\parabf{Internal Validity.} One potential threat to internal validity is the selection of hyperparameters, which can significantly impact model performance. To address this, we followed the hyperparameter settings recommended in \cite{behnamghader2024llm2vec} for all model fine-tuning, ensuring consistency and reliability in our evaluations.
Furthermore, the pretraining datasets for decoder-only LLMs may contain CSN data, which could undermine the reliability of model evaluations. To mitigate this concern, we also tested the models on a more recent dataset, \cosqaplus{}, ensuring that the models had not been exposed to the relevant knowledge during pretraining. 
This additional measure helps ensure that the evaluation results reflect the model's true capabilities rather than any bias introduced by prior exposure to the dataset.

\parabf{External Validity.}
The study used \revised{eleven} decoder-only LLMs, which might limit generalizability. To address this, we selected models that rank high in code generation on MBPP \cite{paperswithcode2024} and have similar sizes, ensuring that our models are SOTA and representative. We prioritized using official models, and if unavailable, we chose those with high downloads and comprehensive evaluations to ensure model reliability. 
\revRtwo{
In the analysis of RQ3, different aspects face distinct validity threats. For \ref{sec:train-Mothod}, our qualitative analysis presents results primarily for the Go language. This might limit the generalizability of these specific visual findings. However, we mitigate this by performing the same analysis on all other CSN languages and confirming that they all exhibit similar trends.
For \ref{Training_Data} and \ref{sec:rq3_Model_Size}, our analyses are limited to the selected datasets and model families. To mitigate this threat, we employ rigorous statistical tests to ensure our findings are statistically significant.
}
\revised{Furthermore, the analyses for RQ4 and RQ5 were conducted exclusively on CodeGemma, our top-performing 7B model. This single-model analysis limits the generalizability of our findings on computational time and training efficiency, as these metrics are sensitive to model-specific factors like architecture, pre-training data, and hyperparameters.}

%% file: sections/Practical_Implications.tex
\section{discussion}
\revRtwo{In the session, we discuss some practical implications for researchers and developers. In addition, we discuss potential research directions for future work.}

\subsection{\revised{Practical Implications}}

\revised{Based on our systematic evaluation, we offer the following recommendations for those looking to apply language models to code search tasks:}

\revised{\textbf{For Zero-Shot Scenarios}: When fine-tuning is not feasible due to resource or time constraints, we recommend using specialized encoder-only models like UniXcoder. Despite its smaller size, it offers a competitive balance of performance and efficiency, outperforming many recent, larger decoder-only LLMs in zero-shot settings.}

\revised{\textbf{For Fine-Tuning Scenarios}: If resources for fine-tuning are available, we strongly recommend decoder-only LLMs, particularly models from the Gemma \revRtwo{and Qwen2.5-Coder} \revRtwo{families}. Our results show that \revRtwo{their architectures demonstrate} superior understanding and generalization capabilities after fine-tuning, surpassing even \revRtwo{other} recent and larger models.}

\revised{\textbf{For Fine-tuning Method}: We recommend using supervised contrastive learning (SupCon). This method is highly effective at improving a model's ability to distinguish between relevant and irrelevant code snippets, which is the core of the code search task.}

\revised{\textbf{For Training Data:} Fine-tuning on a specialized code search dataset yields statistically significant improvements over using a general-purpose information retrieval dataset. Training on a multilingual dataset can bring stronger generalization ability and more stable performance compared to using a single-language dataset. Meanwhile, the data composition is critical, as a small amount of data from a specific language may interfere with the model's training effectiveness, potentially acting as statistical noise.}

\revised{\textbf{For Model \revRtwo{Size}}: Model size is not always proportional to performance. Our experiments indicate that a larger model does not guarantee better results for code search. We \revRtwo{observe} a non-monotonic trend where a mid-sized model (e.g., 1.5B) sometimes outperforms its larger counterparts. This suggests that practitioners may find a better cost-performance balance with moderately-sized models, but should determine the optimal size through targeted experiments.}

\revised{\textbf{For Resource Trade-offs:} The cost of fine-tuning should be justified. While the offline cost of fine-tuning a large model is significant, it is a one-time investment that can yield a substantial and lasting improvement in the final system's performance.}

\subsection{Future Work}\label{futureWork}

\revRtwo{Our findings not only provide practical guidance but also open up several avenues for future research:}

\begin{itemize} 
\item \revRtwo{\textbf{Developing Next-Generation Code Search Datasets:} Our study confirms that fine-tuning on specialized code search data (like CSN, which uses code comments as queries) yields superior results compared to general-purpose text. However, the nature of code search is evolving. Modern applications, such as RAG and AI agents, will generate queries that are far more diverse than traditional human-authored comments. This indicates a critical need for new, high-quality, and varied fine-tuning datasets that reflect both complex human intent and AI-generated retrieval queries.}

\item \revRtwo{\textbf{Accelerating Fine-Tuning without Performance Loss:} Although the current cost of full fine-tuning is a one-time investment, it remains substantial. For decoder-only LLMs, increasing the fine-tuning speed while ensuring no degradation in performance would significantly enhance their practical appeal. Future work should systematically evaluate the trade-offs of PEFT methods specifically for the code search task, aiming to optimize both training efficiency and final retrieval accuracy.}

\item \revRtwo{\textbf{Investigating the Non-Monotonic Scaling Effect:} Our discovery that model size does not monotonically improve performance (with a 1.5B parameter model outperforming larger counterparts) warrants deeper investigation. Future research should explore why this occurs. Is it due to optimization difficulties, the nature of the code search task, or catastrophic forgetting during fine-tuning? Identifying the ``sweet spot" for model scale in code search could lead to more efficient and effective models.}

\item \revRtwo{\textbf{Optimizing Data Composition:} We note that small amounts of data from specific languages might act as ``statistical noise" and hinder training. Future work could quantify this effect more precisely. What is the minimum data threshold required for a language to contribute positively to a multilingual model? Can curriculum learning or data re-weighting schemes mitigate the negative impact of low-resource languages in a multilingual dataset?}

\end{itemize}

%% file: sections/conclusion.tex
\section{Conclusion}
This study evaluates \revised{eleven} state-of-the-art decoder-only LLMs for code search tasks. While these models initially underperform in zero-shot settings due to mismatched code representations, fine-tuning significantly boosts their performance, allowing them to better leverage pre-trained code understanding. Fine-tuned CodeGemma , \revRtwo{for instance, achieves SOTA performance on CSN and top-tier performance on \cosqaplus, underscoring the importance of specialized pre-training.}

Our analysis shows that fine-tuning on code-specific datasets, utilizing supervised contrastive learning, and mid-sized models contribute to performance improvements. However, model architecture remains critical, as larger models do not always guarantee superior results. Decoder-only models excel in long-code searches but struggle with ultra-short queries due to the curse of dimensionality and insufficient context. Although the larger size of these models leads to longer computational times, the costs remain manageable. Moreover, decoder-only LLMs demonstrate superior training efficiency and generalization on limited data compared to smaller encoder-only models.

In summary, this study highlights the potential of fine-tuned decoder-only LLMs for code search tasks, demonstrating enhanced performance and generalization. \revRtwo{Our findings offer a comprehensive guide for practitioners on model selection, data strategy, and performance trade-offs, while also highlighting key avenues for future research, such as developing next-generation datasets and investigating the non-monotonic scaling effects we identified.}